\documentclass[prc,preprint,showpacs,showkeys]{revtex4}

\usepackage{graphics}
\usepackage{graphicx}

\newcommand{\be}{\begin{equation}}
\newcommand{\ee}[1]{\label{#1} \end{equation}}
\newcommand{\ba}{\begin{eqnarray}}
\newcommand{\ea}[1]{\label{#1} \end{eqnarray}}

\newcommand{\rep}{ {\Re e \, P} }

\newcommand{\vs}{\vspace{0mm}}
\newcommand{\exv}[1]{ {\left\langle {#1} \right\rangle} }

\begin{document}

\title{{\bf Polyakov Loop Behavior in Non-Extensive SU(2) Lattice Gauge Theory}}

\author{ 
  T. S. Bir\'o
}

\affiliation{
  {KFKI Research Institute for Particle and Nuclear Physics,} \\
  { H-1525 Budapest, P.O.Box 49, Hungary} 
}

\author{
 Z. Schram
}

\affiliation{
{Department for Theoretical Physics, University of Debrecen,} \\
  { H-4010 Debrecen, P.O.Box 5, Hungary}
}

\keywords{non-extensive thermodynamics, lattice gauge theory, Polyakov Loop}
\pacs{65.40.gd, 25.75.Ag}

\date{\today}

\begin{abstract}
  In order to come closer to a realistic model of high-energy collisions,
  we simulate SU(2) lattice gauge theory under fluctuating
  temperature. The fluctuations are Euler-Gamma distributed,
  leading to a canonical state maximizing the R\'enyi and Tsallis
  entropy formulas. This choice conforms to the multiplicity distributions
  leading to the KNO scaling in high energy experimental spectra.
  We test the random lattice spacing method
  numerically by investigating the Polyakov Loop expectation value,
  known to be a good order parameter for the confinement -- deconfinement
  phase transition in ordinary canonical Monte Carlo methods.
  The critical coupling (and presumably the temperature) move with the 
  width parameter of the inverse temperature fluctuations towards higher values.
\end{abstract}

\maketitle
\normalsize

\section{Introduction}

Lattice gauge theory is up to now the only successful nonpertubative numerical approach
to solve physical problems related to the strong interaction. Among the most reknown
recent results the prediction of a critical endpoint of the phase transition in QCD
became in the forefront of research\cite{FODOR1,FODOR2,PETREC1,PETREC2}. 
Also a large scale experimental program,
FAIR at GSI, has been initiated, among other goals for studying the interface between quark-
and hadronic matter in the CBM experiment \cite{CBM}.
Accelerator experiments, however, do not have a control on thermodynamically 
relevant parameters, like the temperature and pressure, to such a degree that these
could be regarded as having a sharp and constant value during the evolution of the strongly
interacting matter. Lattice theoretical simulations on the other hand assume a fixed
value for the temperature.

Our aim with the study presented in this paper is to move towards a more flexible scheme:
we treat temperature as a random variable, defined not only by its expectation value, but also
by a width. In fact the thermodynamically consequent approach to this problem requires
that the inverse temperature, $\beta=1/k_BT$, occurring also as a Lagrange multiplier
for the fixed energy constraint by maximizing the entropy, is fixed on the average and then
randomized. Such a superstatistical method \cite{SUPER1,SUPER2,SUPER3,SUPER4,SUPER5,SUPER6} 
is in accord with recent findings on non-extensive thermodynamics, 
where the canonical energy distribution is not-exponential, but rather shows an experimentally 
observed power-law tail \cite{POWER1,POWER2,POWER3,POWER4,POWER5,POWER6}.

In this paper we review basic thermodynamic arguments to relate the temperature to the parameters
of a statistical power-law tailed, canonical energy distribution. Following this the superstatistical
method is presented, in particular its realization strategy for lattice Monte Carlo simulations.
We choose to randomize the timelike to spacelike lattice spacing ratio, $\theta=a_t/a_s$.
The most important first task is to check the deconfinement phase transition by observing
the Polyakov loop expectation value. These results are presented and discussed.
As a main consequence we predict that the deconfinement transition temperature is likely to be higher
than determined by fixed-$T$ lattice calculations so far.

\section{Thermodynamical Background}


Based on arguments regarding the compatibility of general composition rules
for the total entropy and energy of composed thermodynamical systems \cite{POWER6},
in an extended canonical thermal equilibrium problem the absolute temperature
is given by
\be
 \beta = 1/T = \partial \hat{L}(S)/\partial L(E),
\ee{ABSOLUT_THERM_TEM}
with $\hat{L}(S)$ and $L(E)$ being the additive formal logarithms of the
respective composition formulas. The formal logarithm maps a general composition law,
say $S_{12}=S_1\oplus S_2$, to the addition by $\hat{L}(S_{12})=\hat{L}(S_1)+\hat{L}(S_2)$.
This construction leads us to maximize
$\hat{L}(S)-\beta L(E)$ when looking for canonical energy distributions \cite{POWER5}.
The probability distribution, $w_i$,  of states with energy $E_i$ in equilibrium 
maximizes the formal logarithm of the non-extensive entropy formula with constraints
on the average value of the also non-additive energy and the probability normalization:
\be
  \hat{L}(S)\left[ w_i \right] - \beta \sum_i w_i L(E_i) -\alpha \sum_i w_i = {\rm max}.
\ee{GENERAL_CANONICAL}
Here $\beta$ and $\alpha$ are Lagrange multipliers and it can be proven that
$\beta=1/T$ is related to the thermodynamically valid temperature according to
the zeroth law of thermodynamics.
Choosing the next to simplest composition formula to the addition, supplemented with a
leading second order correction,
\be
 S_{12}=S_1+S_2+\hat{a}S_1S_2,
\ee{ENTROPY_TSALLIS_COMPO}
the additive formal logarithm function is given by
\be
 \hat{L}(S)=\frac{1}{\hat{a}} \ln(1+\hat{a}S).
\ee{TSA_FORM_LOG}
This way $\hat{L}(S_{12})=\hat{L}(S_1)+\hat{L}(S_2)$, indeed.
By using the Tsallis entropy formula \cite{NEXT1,NEXT2,NEXT3,NEXT4,NEXT5},
\be
 S = \frac{1}{\hat{a}}  \sum_i \left( w_i^{1-\hat{a}} - w_i \right),
\ee{TSALLIS_ENTROPY}
this formal logarithm turns out to be the R\'enyi entropy \cite{RENYI1,RENYI2}
\be
\hat{L}(S) = \frac{1}{\hat{a}} \ln \sum_i w_i^{1-\hat{a}}.
\ee{RENYI_ENTROPY}
It is customary to use the parameter, $q=1-\hat{a}$. 
The above power-law tailed form of energy distribution can be fitted to
experimentally observed particle spectra, and this way a numerical value
for the parameter $\hat{a}$ can be obtained.
The $\hat{a}=0$ ($q=1$) case recovers the classical Boltzmann-Gibbs-Shannon (BGS) 
formula \cite{GE1,GE2,GE3,GE4}
\be
 S_{BG} = \sum_i - w_i \ln w_i.
\ee{BGS_ENTROPY}

According to this the quantity \(\hat L(S) \) is to be maximized with constraints. 
Identifying the analogous formal logarithm  for leading order non-additive energy composition, 
$E_{12}=E_1+E_2+aE_1E_2$,  as
\be
L(E) = \frac{1}{a} \ln(1+aE), 
\ee{E_FROM_LOG}
one considers
\be
\frac{1}{\hat{a}} \ln  \sum_i w_i^{1-\hat{a}}  - \beta \sum_i w_i \frac{1}{a} \ln(1+aE_i)
- \alpha \sum_i w_i = {\rm max}.
\ee{RENYI_CANONICAL}
The maximum is achieved by the canonical probability distribution 
\be
  w_i =A\left(b(\alpha + \beta L_i)\right)^{-\frac{1}{\hat a}}
\ee{edist}
with
\be
L_i =\frac{1}{a} \ln(1+aE_i), \qquad A = e^{-\hat L(S)}, \qquad b=\frac{\hat a}{1-\hat a}.
\ee{LI}
Then the normalization, the average and  the definition of the entropy 
lead to the condition
\be
1=b\alpha+b\beta\left<L\right>. 
\ee{maincond} 
Finally the equilibrium distribution simplifies to 
\be
 w_i  
 = \frac{1}{Z} \left(1+\hat{a}\hat\beta L_i \right)^{-1/\hat{a}} 
\ee{RENYI_DISTRIB}
with $L_i$ given in eq.(\ref{LI}).
Here we have introduced the following shorthand notations: 
\be
Z = \frac{1}{A } \, (1-b \beta \left<L\right> )^\frac{1}{\hat a}, \qquad 
\hat{\beta} = \frac{\beta}{1-\hat{a}(1+\beta \left<L\right>)}.
\ee{cons1}
We should keep in mind that the reciprocal temperature, distinguished by the Zeroth Law, 
is the Lagrange multiplier \(\beta\). This is reflected well by the whole formalism, 
because the usual thermodynamic relations are valid. 

It is particularly interesting to consider now cases, when only one of the two quantities is
composed by non-additive rules.
In the limit of additive entropy but non-additive energy 
(\(\hat a \rightarrow 0 \)) the canonical distribution approaches
\be
 w_i = \frac{1}{Z_{0}} \left( 1+aE_i\right)^{-\beta/a},\qquad
 \text{where} \qquad \ln  Z_0 = S_{BG} - \beta \left<E\right>. 
\ee{ENTR_ADD_ENERG_NON}
Here \(S_{BG}\) is the Boltzmann-Gibbs-Shannon entropy (cf. eq.\ref{BGS_ENTROPY}). 
For non-additive entropy and additive energy on the other hand a similar, but differently parametrized
power-law tailed distribution emerges:
\be
 w_i = \frac{1}{Z} \left( 1+\hat{\beta} \hat{a} E_i\right)^{-1/\hat{a}},
\ee{ENERG_ADD_ENT_NON}
with
\be
\hat{\beta} = \frac{\beta}{1-\hat{a}(1+\beta\exv{E}) }.
\ee{HAT_BETA}
The latter relation can be transformed into a more suggestive form by using $q=1-\hat{a}$ and the
temperature parameters $T=1/\beta$ and $\hat{T}=1/\hat{\beta}$:
\be
T = \frac{1}{q} \hat{T} + \left(\frac{1}{q}-1 \right) \exv{E}.
\ee{HAT_T_T}
By using the distribution given in eq.(\ref{ENERG_ADD_ENT_NON}), the expectation value of the
energy, $\exv{E}$, is directly given as a function of $\hat{T}$ and $\hat{a}=1-q$. 


\section{Superstatistical Monte Carlo Method}

%
%
%

\vs

\vs
In either case discussed in the previous section, the generalized canonical
distribution of the different energy states in a system in thermal equilibrium
with non-additive composition rules is given by a formula
\be
 w_i = \frac{1}{Z_{TS}} \: \left( 1 + \frac{\beta E_i}{c} \right)^{-c}.
\ee{POWER_WEIGHT}
In the $c \rightarrow\infty$ limit this formula coincides with the
familiar Gibbs factor:
\be
 \lim_{c\rightarrow\infty} w_i = \frac{1}{Z_G} \exp(-\beta E_i).
\ee{GIBBS_LIMIT}
The quantity $q=1-1/c$ is called the Tsallis index. 
Here $c=\beta/a$ and $\beta$ is in fact the inverse absolute temperature for the energy non-additivity case;
for the entropy non-additivity on the other hand $\beta$ has to be replaced by $\hat{\beta}$ and $c$ by $1/\hat{a}$ 
as it was explained in the previous section.
The thermodynamic temperature in the latter case, according to the Zeroth Law, 
can be obtained by using eq.(\ref{HAT_T_T}).

\vs
The Tsallis distribution weight factor, $w_i$, 
on the other hand can be obtained as an integral
of Gibbs factors over the Gamma distribution \cite{NEXLAT1,NEXLAT2},
\be
 w_i = \frac{1}{Z_{TS}} \, \int_0^{\infty}\!d\theta \, 
 w_c(\theta) \exp(- \theta \beta E_i), 
 \ee{SUPER_WEIGHT}
with
\be
 w_c(\theta) = \frac{c^c}{\Gamma(c)} \, \theta^{c-1} \, e^{-c\theta}.
\ee{NORM-GAMMA}
$\Gamma(c)=(c-1)!$ for integer $c$ is Euler's Gamma function. By its definition the integral of
$w_c(\theta)$ is normalized to one. This approach is a particular case of the so called {\em superstatistics} 
\cite{SUPER1,SUPER3}.

\vs
Based on this, any canonical Gibbs expectation value, if known as a function
of $\beta$, can be converted into the corresponding expectation values
with the power-law tailed canonical energy distribution.
The respective partition functions, $Z_G$ and $Z_{TS}$ ensure the normalization
of the $w_i$ probabilities, $\sum_i w_i = 1$. They are related to each other:
\be
 Z_{TS}(\beta) = \sum_i \int_0^{\infty}\!d\theta \, w_c(\theta) 
 \exp(-\theta \beta E_i) = 
  \int_0^{\infty}\!d\theta \, w_c(\theta)  Z_G(\theta \beta).
\ee{ZTS}
The above formula can be interpreted as averaging over different $\theta \beta$-valued 
Gibbs simulations. The averaging is understood in the partition sum, meaning
that the weighting 'Boltzmann'-factor is also fluctuating. It assumes that the underlying process of
mixing different inverse temperatures is much faster than the averaging itself.

\vs

The question arises, which strategy is the best to follow in order
to perform lattice field theory
simulations with power-law tailed statistics instead of the Gibbs one.
Neither the ensemble of different $\beta$ values (Euclidean timelike lattice
sizes), nor the re-sampling of the traditional, Gibbs distributed
configurations is practicable in a naive way. The $N_t$ lattice sizes are
limited to a small number of integer values -- hence the good coverage
of a Gamma distribution with an arbitrary real $c$ value is questionable.
The already produced configuration ensembles were selected by a Monte
Carlo process according to the Gibbs distribution with the original
lattice action; there is no guarantee that the re-weighting procedure
(which includes part of the weight factors in the operator expressions
for observables) is really convergent (i.e. does not contain
parts growing exponentially or worse). 
We choose another strategy: we use $\theta$ values selceted as random deviates
from an Euler-Gamma distribution during the Monte Carlo statistics.

\vs
The lattice simulation incorporates the physical temperature by the period length in the
Euclidean time direction: $\beta = N_ta_t$. Due to the restriction to a
few integer values of $N_t$, 
we simulate the Gamma distribution of the physical $\beta=1/T$ values by
a Gamma distribution of the timelike link lengths, $a_t$.
We assume that its mean value is equal to the spacelike lattice spacing, $a_s$. 
Then the ratio $\theta=a_t/a_s$ follows
a normalized Gamma distribution with the mean value $1$ and a width of
$1/\sqrt{c}$. (In the view of ZEUS $e^+e^-$ data $c \approx 5.8 \pm 0.5$,
the width is about $40$ per cent.) In our numerical calculations we
apply the value $c=5.5$.

\vs
For calculating expectation values in field theory a generating functional
based on the Legendre transform of $Z$ is used. 
Our starting assumption is the formula (\ref{ZTS})
with
\be
 Z_G\left[\theta \beta\right] = \int {\cal D}U \, e^{-S\left[U,\theta \right]}.
\ee{STARTING}
Since we simulate the
canonical power-law distribution by a lattice with fluctuating
asymmetry ratio, there are two limiting strategies to execute the
Legendre transformation: i) in the {\em annealing} scenario the lattice
fluctuates slowly and one considers first summations over field configurations,
in the ii) {\em quenched} scenario on the contrary, the lattice fluctuations
are fast, form an effective action (virtually re-weighting the occurrence
probability of a field configuration), and the summation over possible
field configuration is the slower process performing the second 
(i.e. the path-) integral. In this paper we investigate numerically the general case
when one may choose when a new value for $\theta$ is taken. The frequency of
these fluctuations may go from one in each Metropolis step for the field
configurations to one in the whole Monte Carlo process (the latter being the
traditional method). Our results presented in the next section belong to
a choice of $5$ field updates for the whole lattice before choosing a new $\theta$.
This peculiar value was controlled by a series of simulations and proved to be
sufficient for a close equilibration to a given, momentary temperature \cite{ACTADEB}.

\vs
The effect of $\theta$ fluctuation is an effective weight
for field configurations, which may depend on a scaling power according to
the time (or energy) dimension of the operator under study. In general we
consider
the Tsallis expectation value of an observable $\hat{A}[U]$ over  lattice field
configurations $U$. $\hat{A}$ may include
the timelike link length, say  with the power $v$:
$\hat{A}=\theta^{\:v}A$.
The Tsallis expectation value then is an average over all possible $a_t$
link lengths according to a Gamma distribution of $\theta=a_t/a_s$.
We obtain:
\be
 \langle A \rangle_{TS} \, =  \, \frac{1}{Z_{TS}} \frac{c^c}{\Gamma(c)}
 \int\!d\theta\: \theta^{\: c-1} e^{-c\:\theta} \int {\cal D}U A\left[U\right]
 \theta^{\:v} e^{-S\left[\theta,U\right]}
\ee{TS-EXP}
with
\be
 Z_{TS} \, =  \, \frac{c^c}{\Gamma(c)}
 \int\!d\theta\: \theta^{\: c-1} e^{-c\:\theta} \int {\cal D}U
 e^{-S\left[\theta,U\right]}.
\ee{Z-TS}
The $\theta$ dependence of the lattice gauge action is known for long:
due to the time derivatives of vector potential in the expression of electric
fields, the ''kinetic'' part scales like $a_ta_s^3/(a_t^2a_s^2)=a_s/a_t$,
and the magnetic (''potential'') part like $a_ta_s^3/(a_s^2a_s^2)=a_t/a_s$
\footnote{This generalizes to all lattice field actions: kinetic and mass
terms scale like $1/\theta$, potential terms like $\theta$.}.
This leads to the following expression for the general lattice action:
\be
 S\left[\theta,U\right] = a \: \theta + b / \theta,
 \label{SLAT}
\ee
where 
$a=S_{ss}[U]$ contains space-space oriented plaquettes and
$b=S_{ts}[U]$ contains time-space oriented plaquettes. The simulation
runs in lattice units anyway, so actually the $U$ configurations are
selected according to weights containing $a$ and $b$. In the 
$c \rightarrow \infty$ limit the scaled Gamma distribution approximates
$\delta(\theta-1)$, (its width narrows extremely, while its integral
is normalized to one), and one gets back the traditional lattice
action $S=a+b$, and the traditional averages.
For finite $c$, one can exchange
the $\theta$ integration and the configuration sum (path integral) and
obtains exactly the power-law-weighted expression. 

\section{Statistics of Polyakov Loops}

\begin{figure}
\begin{center}
	\includegraphics[width=0.44\textwidth]{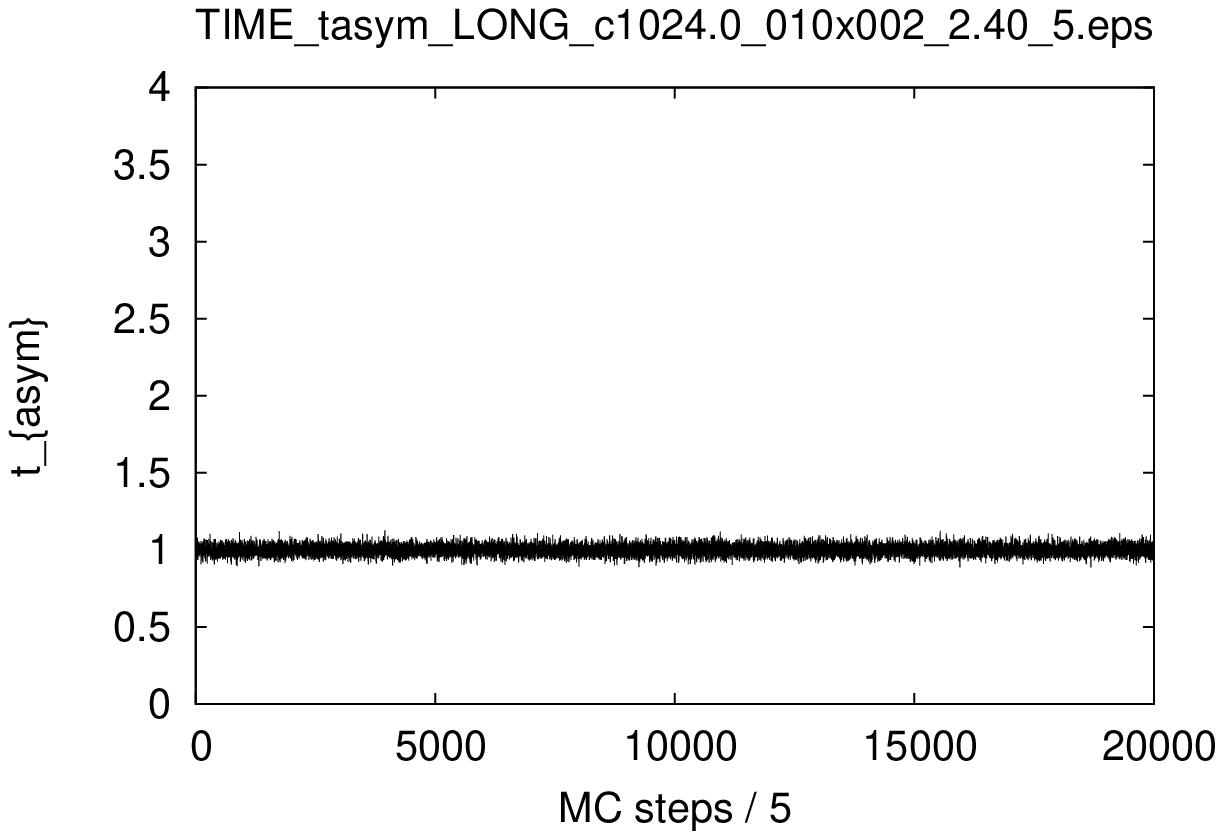} \hspace{4mm} \includegraphics[width=0.44\textwidth]{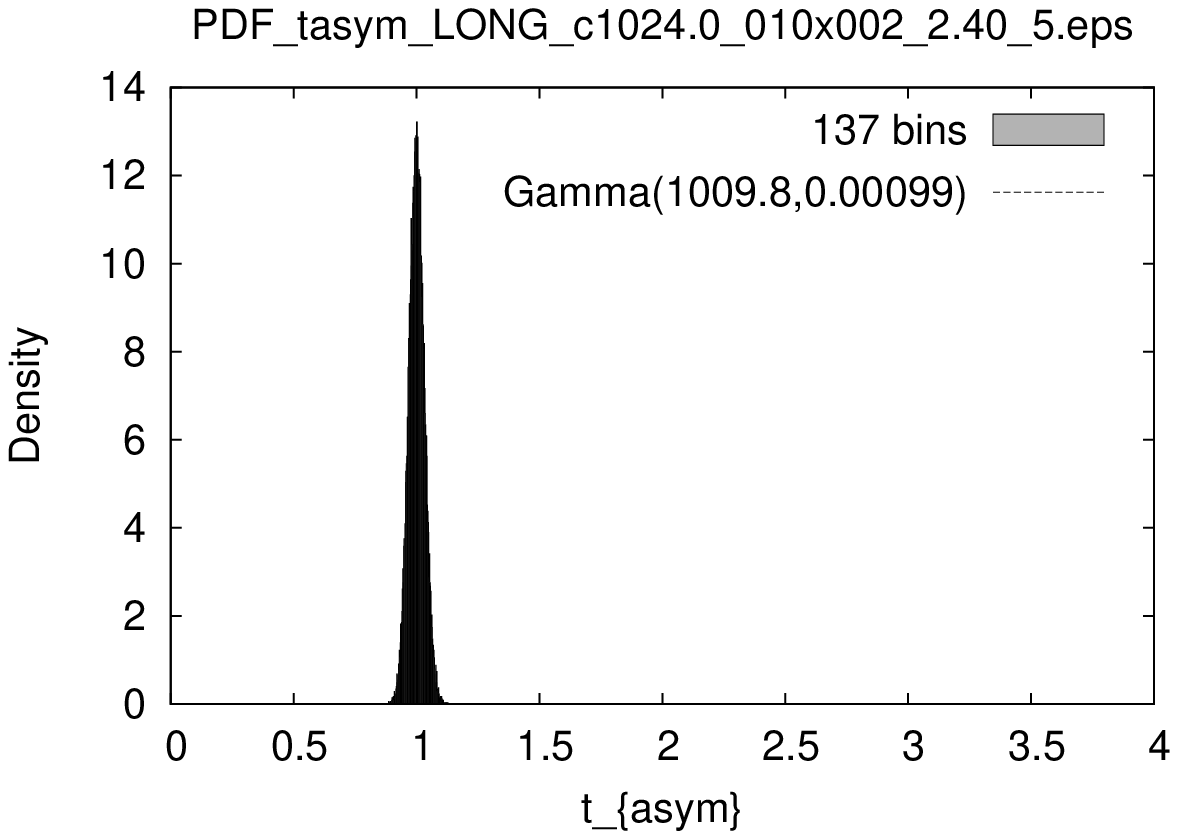}
	\includegraphics[width=0.44\textwidth]{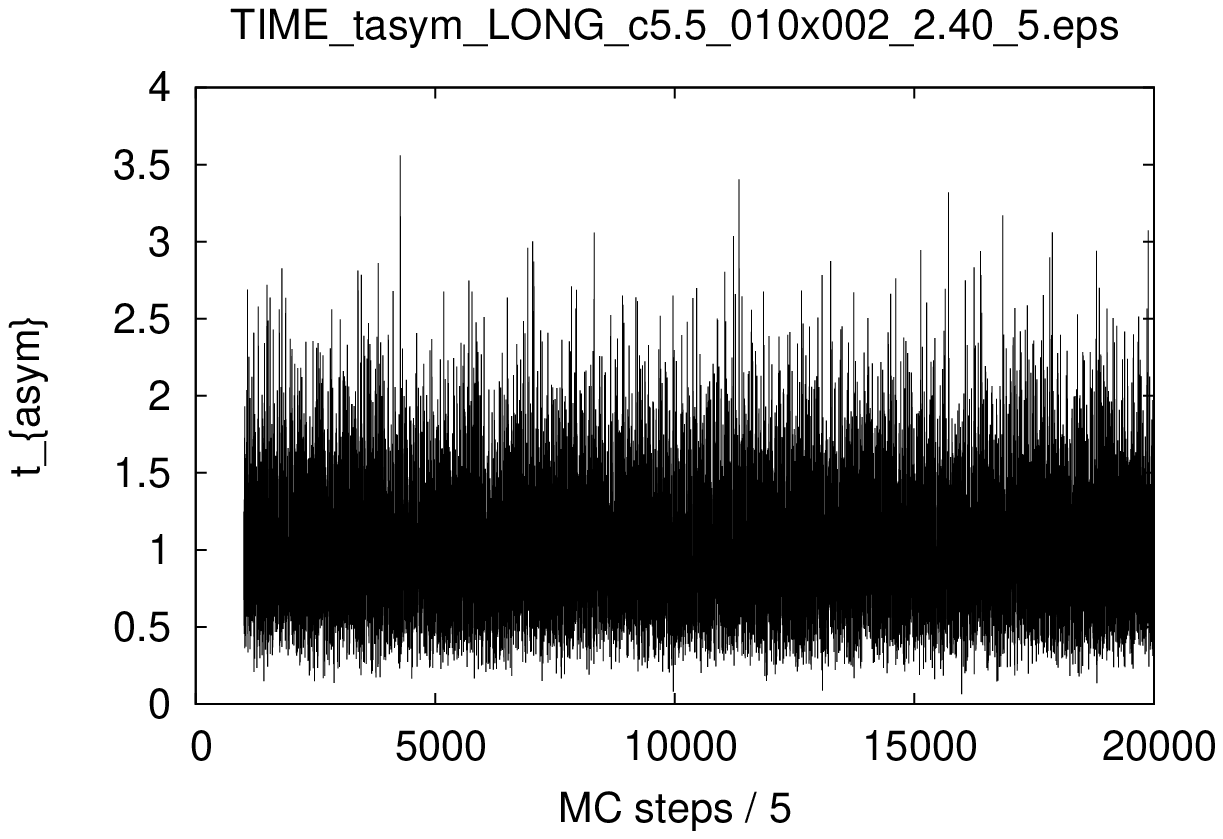} \hspace{4mm} \includegraphics[width=0.44\textwidth]{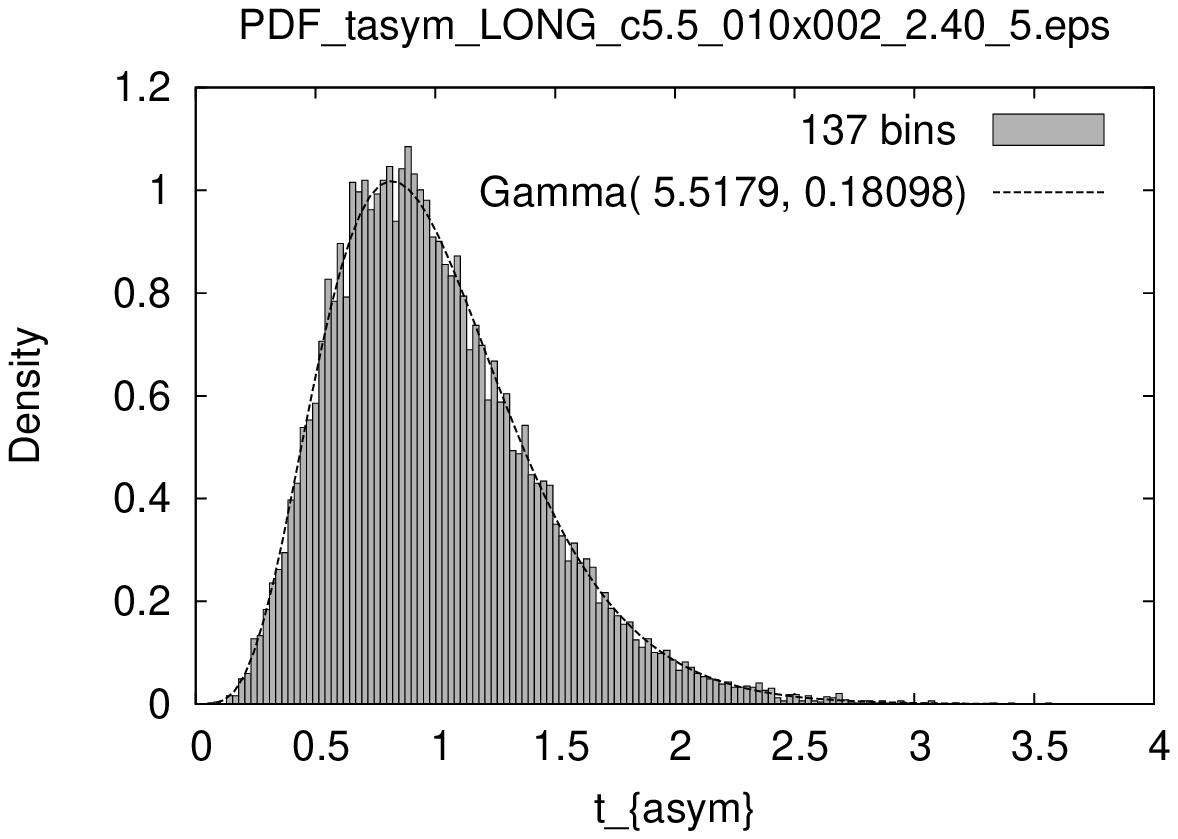}
\end{center}
\caption{ \label{Fig_tasym}
	The Monte Carlo evolution and distribution of $t_{asym}=a_t/a_s$ for the
	coupling $4/g^2 = 2.40$. 
	The random deviates for the results shown in the upper row are thrown with the parameters
	$\alpha = c = 1024.0$ and $\beta = 1/c = 0.000977$. The re-fit by gretl gave
	$\alpha = 1009.8$ and $\beta = 0.000990$.
	The same parameters in the lower row are
	$\alpha = c = 5.5$ and $\beta = 1/c = 0.181818$. The re-fit by gretl gave
	$\alpha = 5.5179$ and $\beta = 0.18098$.
}

\end{figure}

Before discussing our results for the SU(2) pure gauge lattice field simulation
using Euler-Gamma distributed timelike lattice spacing (and simulating this way
a fluctuating inverse temperature to leading order in non-extensive thermodynamics),
let us present a figure about the numerical quality of this randomization.
In Fig.\ref{Fig_tasym} the evolution process and the frequency distribution of
the $\theta$ values are shown for the reference run with $c=1024.0$ and for
the investigated case with $c=5.5$. We have choosen a new value for the asymmetry ratio 
$\theta$ in each 5-th Monte Carlo update -- in order to leave some time for
the relaxation of the field to its thermal state at each instantaneous $\beta\theta$
inverse temperature. In the figure only each $5$-th value is shown. The Monte Carlo
simulations were done at the coupling $4/g^2=2.40$ for this particular statistics
with the Metropolis method.  

Our reference case, thought to be close to the $c=\infty$ traditional system,
is specified by $c=1024$. The re-fit to the distribution of effectively used values
after 20000 draws from the Euler-Gamma distribution by a numerical subroutine
was done by the statistics tool ''gretl''. In the special case of our
random weighting one expects an Euler-Gamma distribution with reciprocial
$\alpha=c$ and $\beta=1/c$ parameters. On the basis of a sample of 20000 $\theta$
values we achieved a reconstruction of $\alpha=1009.8$ and $1/\beta=1010.1$.
Similarly for $c=5.5$ we obtained $\alpha=5.5179$ and $1/\beta=5.5255$.


Now let us turn to the discussion of the behavior of the order parameter of
the confinement -- deconfinement phase transition. The Polyakov Loop 
is calculated by taking the trace of the product of gauge group elements on timelike 
links closing a loop due to the periodic boundary condition:
\be
 P(x) = {\rm Tr} \prod_{t=1}^{N_t}\limits  U_{t}(t,x).
\ee{LOCAL_POLYAKOV_LOOP}
The traditional order parameter of the phase transition is the expectation value 
of the volume averages for each lattice field-configuration during the Monte Carlo
process. For the gauge group $SU(2)$ this quantity is real:
\be
\rep = {\Re e \,} \sum_x P(x). 
\ee{RE_POLYAKOV}
In our present investigations the characteristic width parameter of $1/T$-fluctuations 
is $c=5.5$, corresponding to a relative
width of $1/\sqrt{c} \approx 0.43$. As a reference the $c=1024.0$ case
is taken -- here the relative width is about $1/\sqrt{c} =1/32 \approx 0.03$.

\begin{figure}
\begin{center}
	\includegraphics[width=0.44\textwidth]{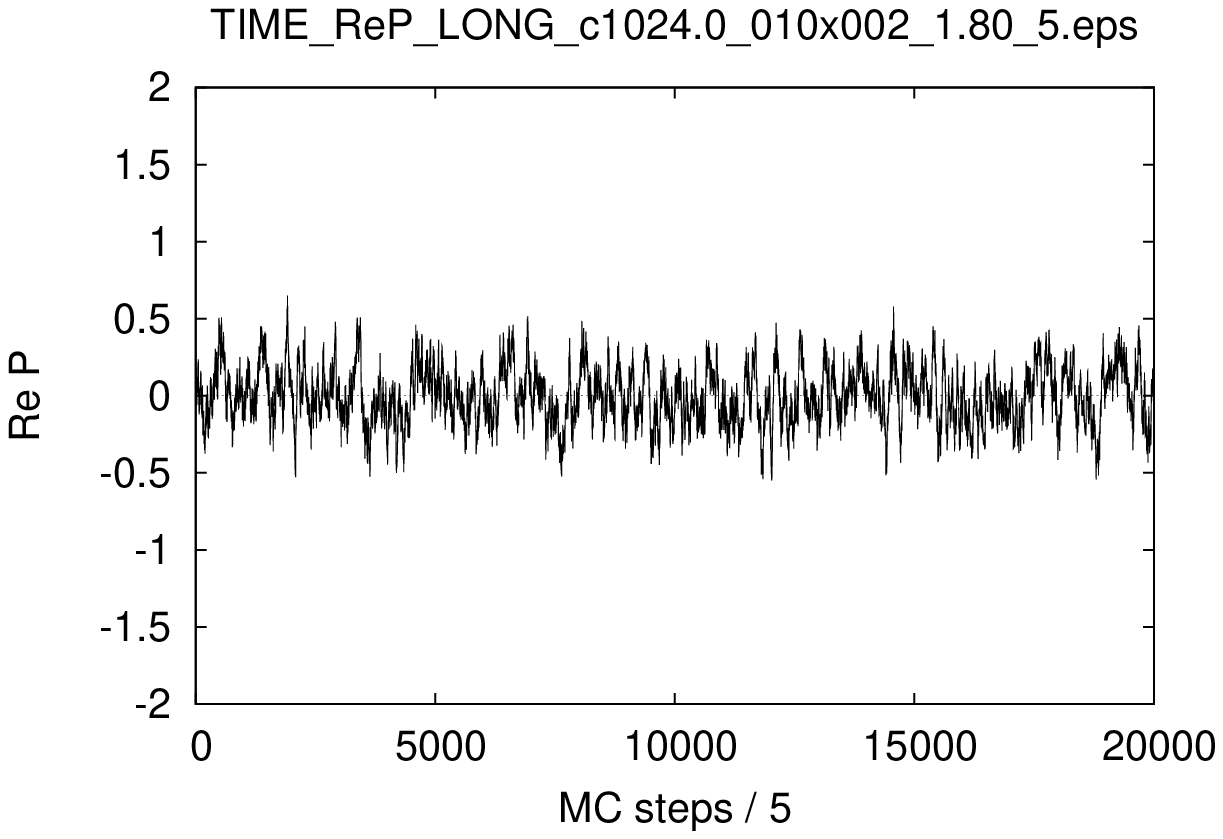} \hspace{4mm} \includegraphics[width=0.44\textwidth]{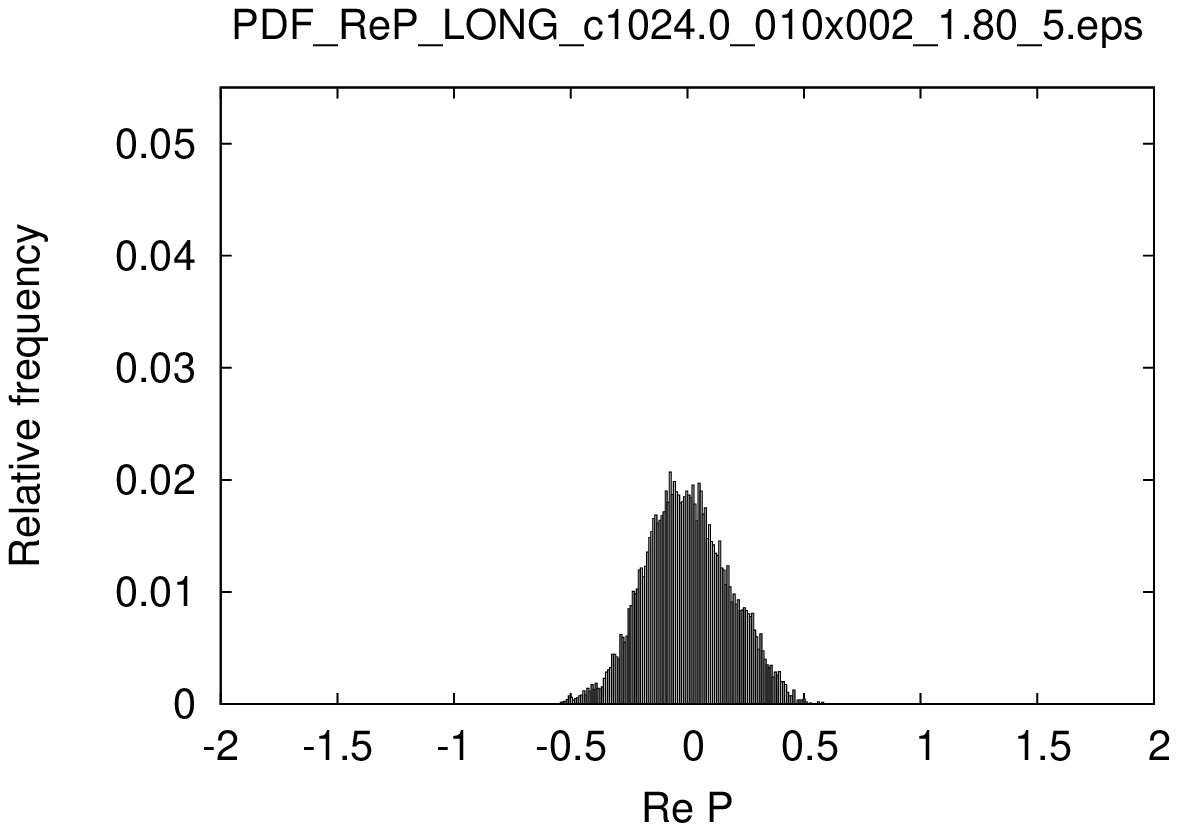}
	\includegraphics[width=0.44\textwidth]{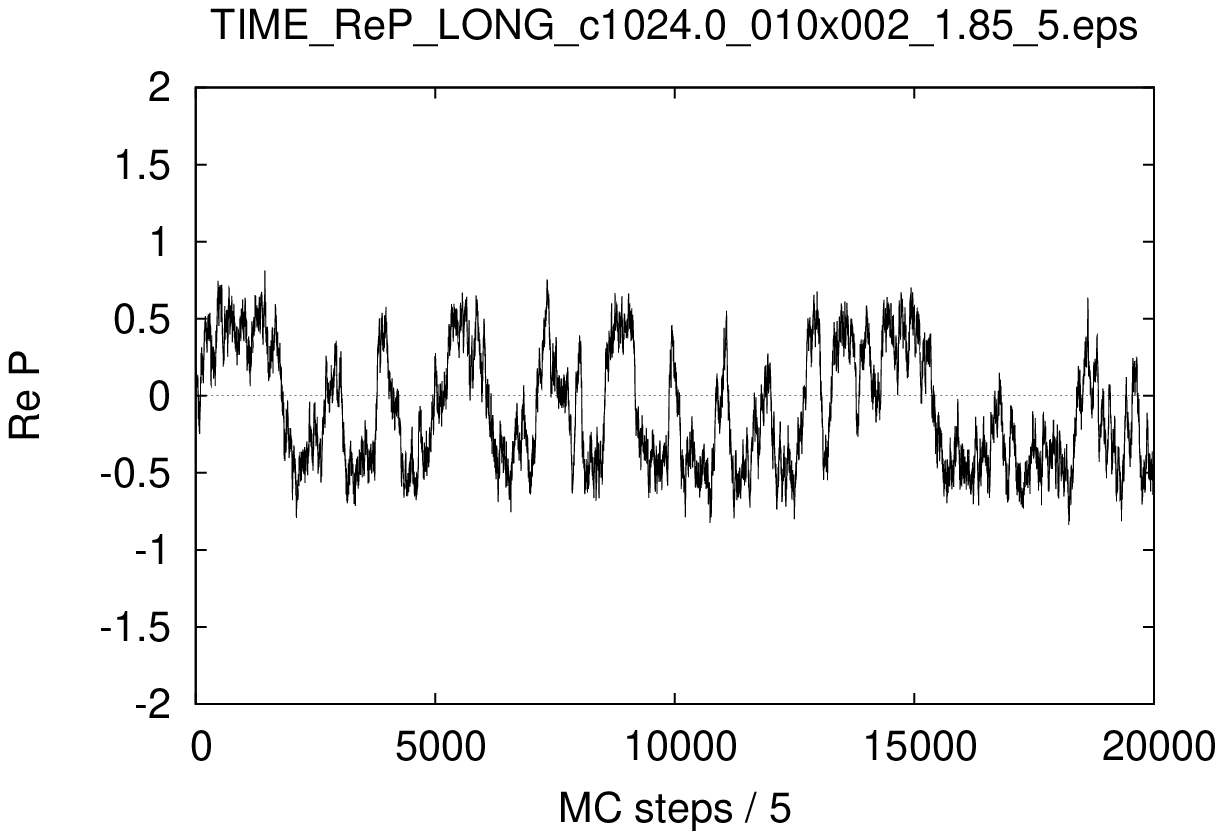} \hspace{4mm} \includegraphics[width=0.44\textwidth]{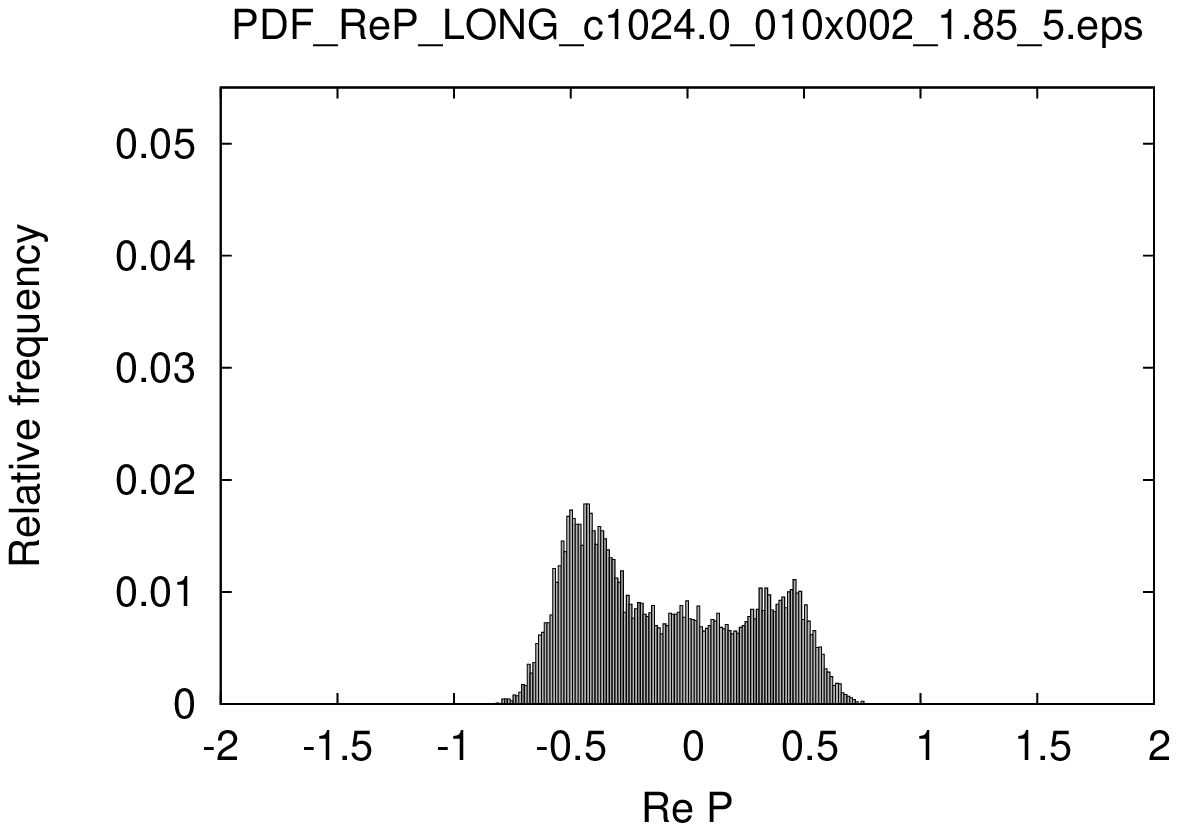}
	\includegraphics[width=0.44\textwidth]{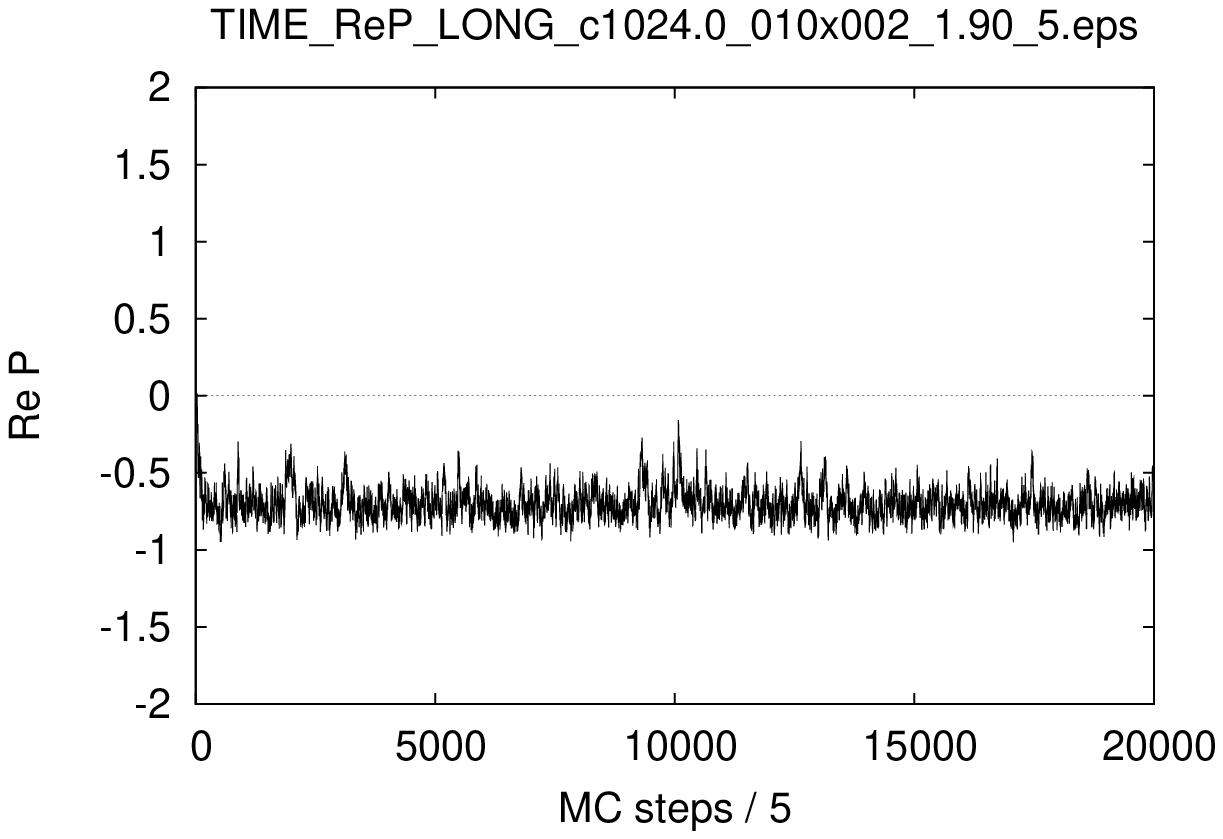} \hspace{4mm} \includegraphics[width=0.44\textwidth]{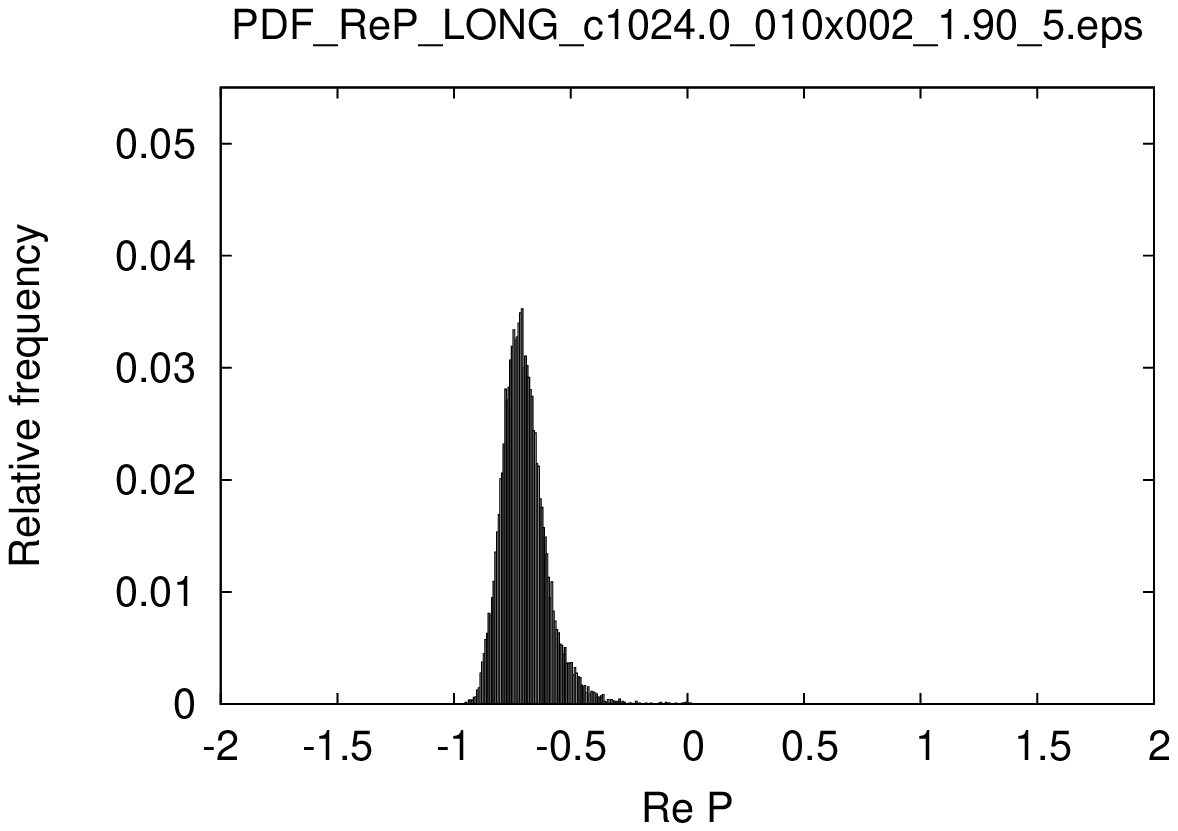}
	\includegraphics[width=0.44\textwidth]{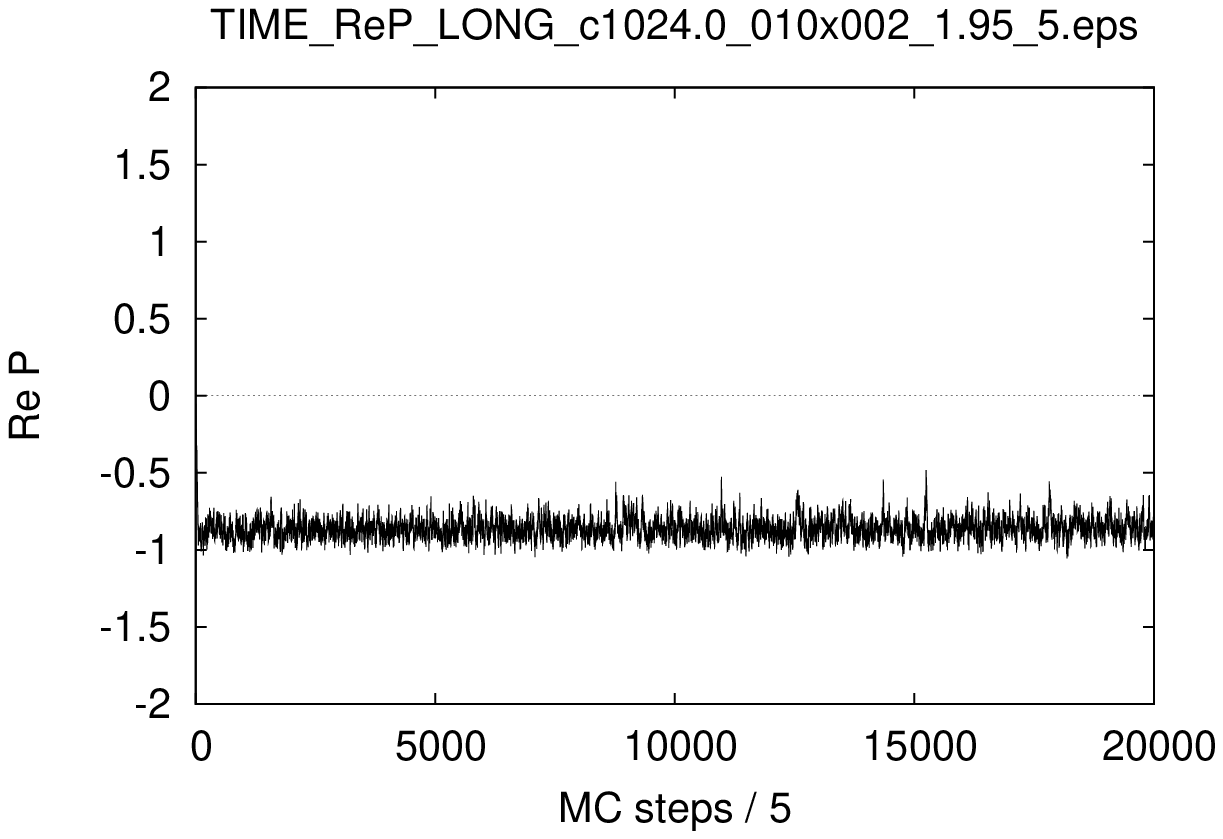} \hspace{4mm} \includegraphics[width=0.44\textwidth]{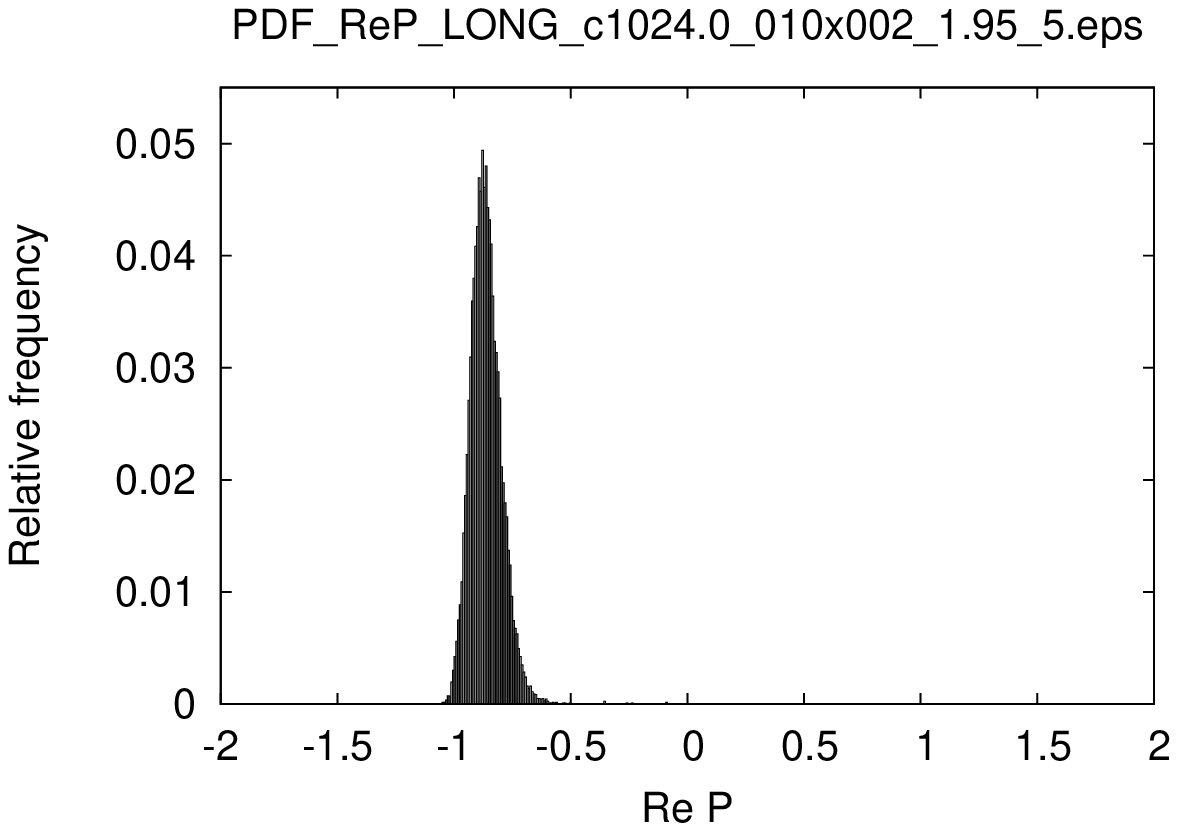}
\end{center}
\caption{ \label{FigReP1024_1.80_1.95}
	The Monte Carlo evolution and distribution of $\rep$ for the
	couplings $4/g^2 = 1.80, 1.85, 1.90$ and $1.95$ using $c=1024.0$ from the top to the bottom.
	This reference pictures show a nearly-traditional confinement -- deconfinement phase transition
	for the SU(2) Yang-Mills system. Note the small width of the order parameter distribution.
}

\end{figure}

The plots in Fig.\ref{FigReP1024_1.80_1.95} show the fluctuations of the order parameter
$\rep$ for the reference runs with $c=1024.0$.
The fluctuating values as a function of the Monte Carlo step are plotted on the left hand side,
while their probability distributions on the right hand side. The values for
the inverse coupling include both the confinement and deconfinement phases.

By producing these results we took five consecutive Metropolis sweeps
over the whole 4-dimensional $10^3\times 2$ lattice while keeping the asymmetry value $\theta=a_t/a_s$ constant.
Then a new $\theta$ was chosen as a random deviate from an Euler-Gamma distribution.
Only these 5-th values are plotted and counted for obtaining expectation values.
The probability distributions of these values were determined by using the statistics
software tool "gretl'. Hereby the first 5000 configurations were sometimes taken
out from the samples, consisting of 100000 lattice configurations each, this did not
change expectation values appreciably. 
For the statistical evaluation  only  each 5-th configuration was selected, being fairly independent
of each other in the evolution governed by the Metropolis algorithm and certainly belonging to different
$\theta$ values.
The frequency distributions reflect cleanly when several $\rep$ expectation
values are occurring during the Monte Carlo evolution, by several maxima. 
This is the case near to the phase transition point.


\begin{figure}
\begin{center}
	\includegraphics[width=0.44\textwidth]{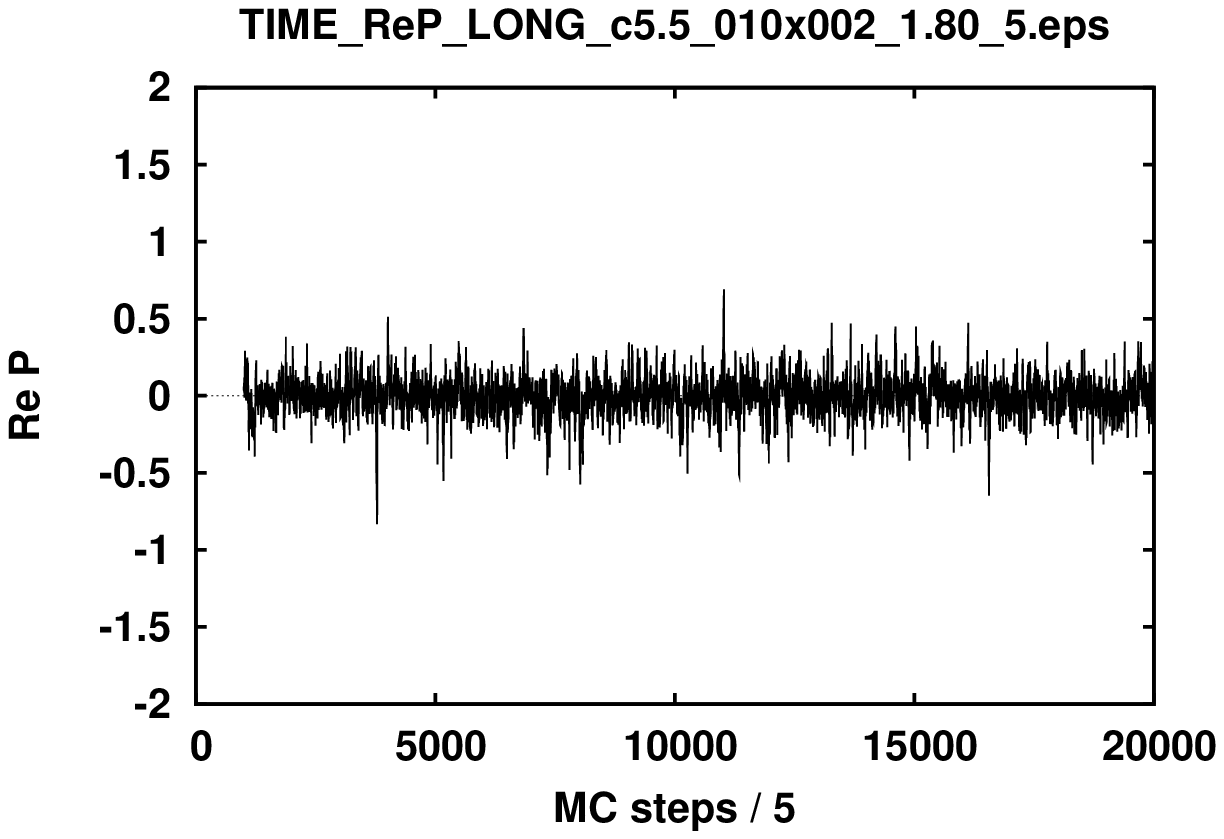} \hspace{4mm} \includegraphics[width=0.44\textwidth]{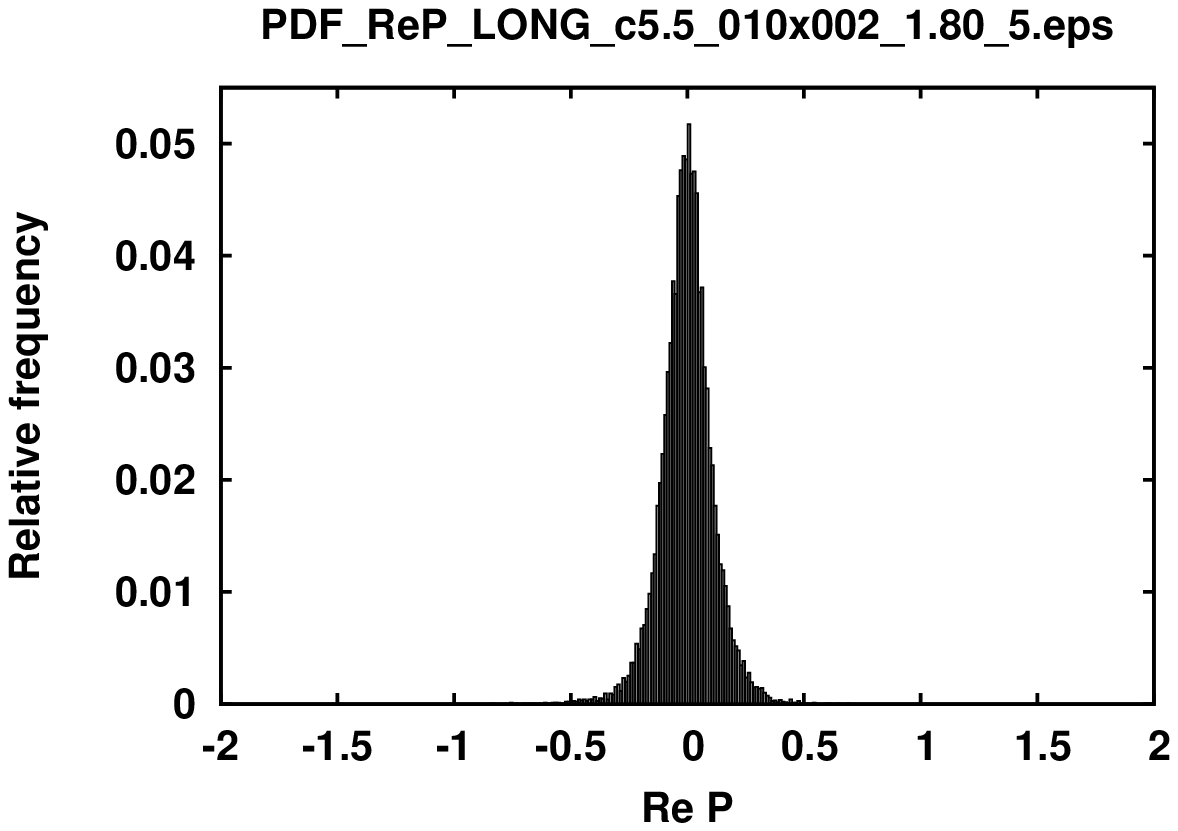}
	\includegraphics[width=0.44\textwidth]{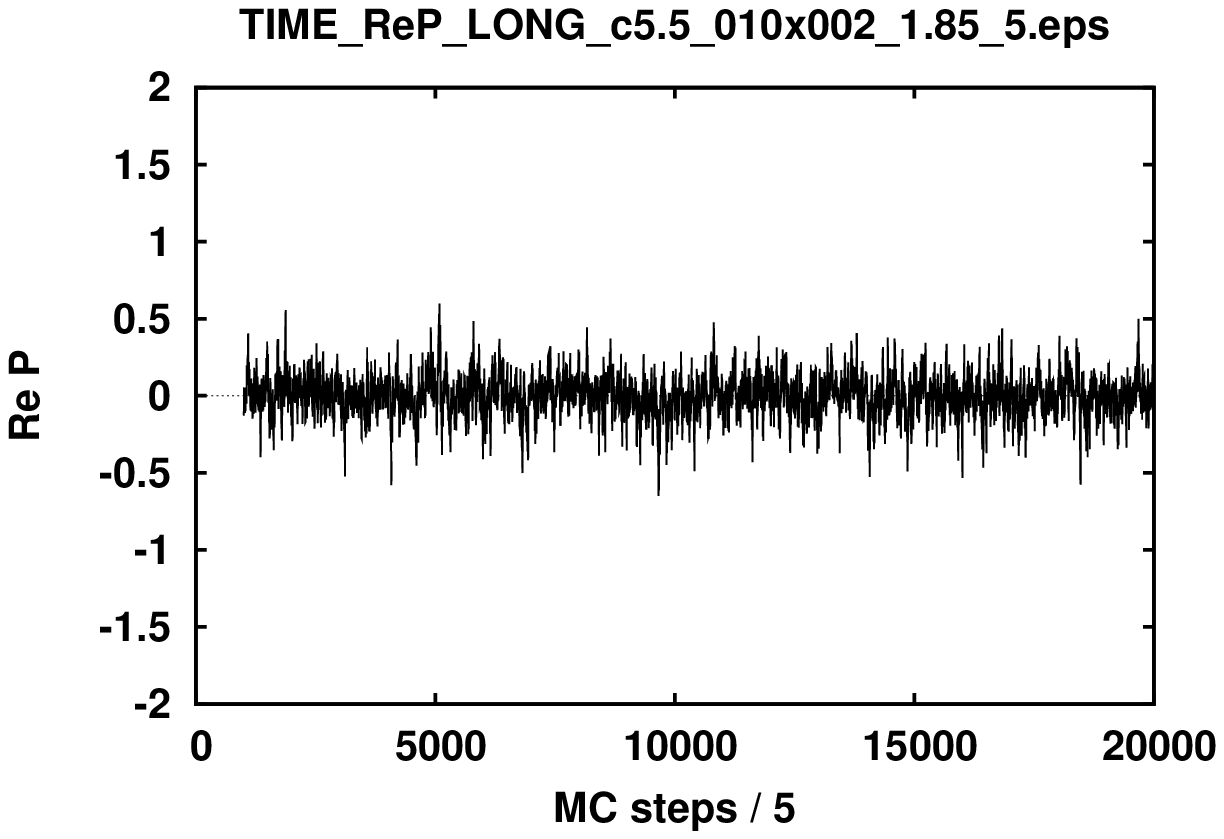} \hspace{4mm} \includegraphics[width=0.44\textwidth]{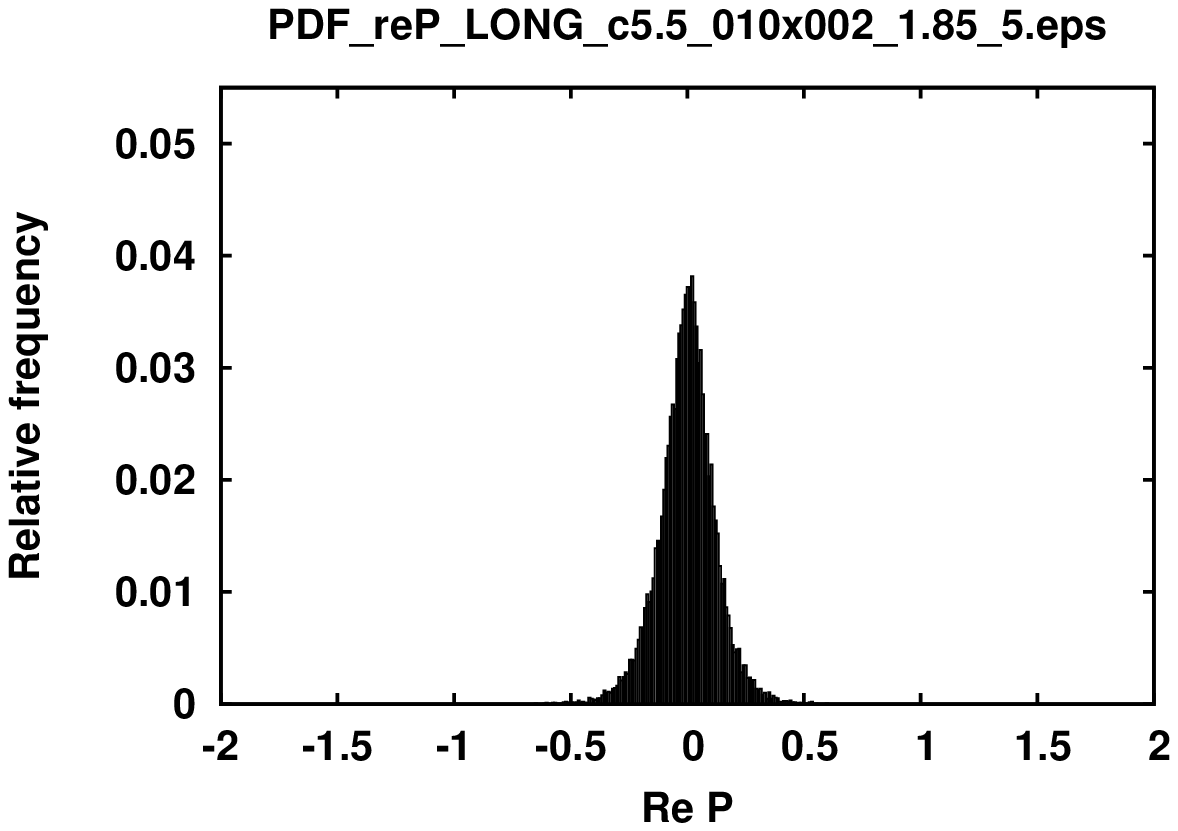}
	\includegraphics[width=0.44\textwidth]{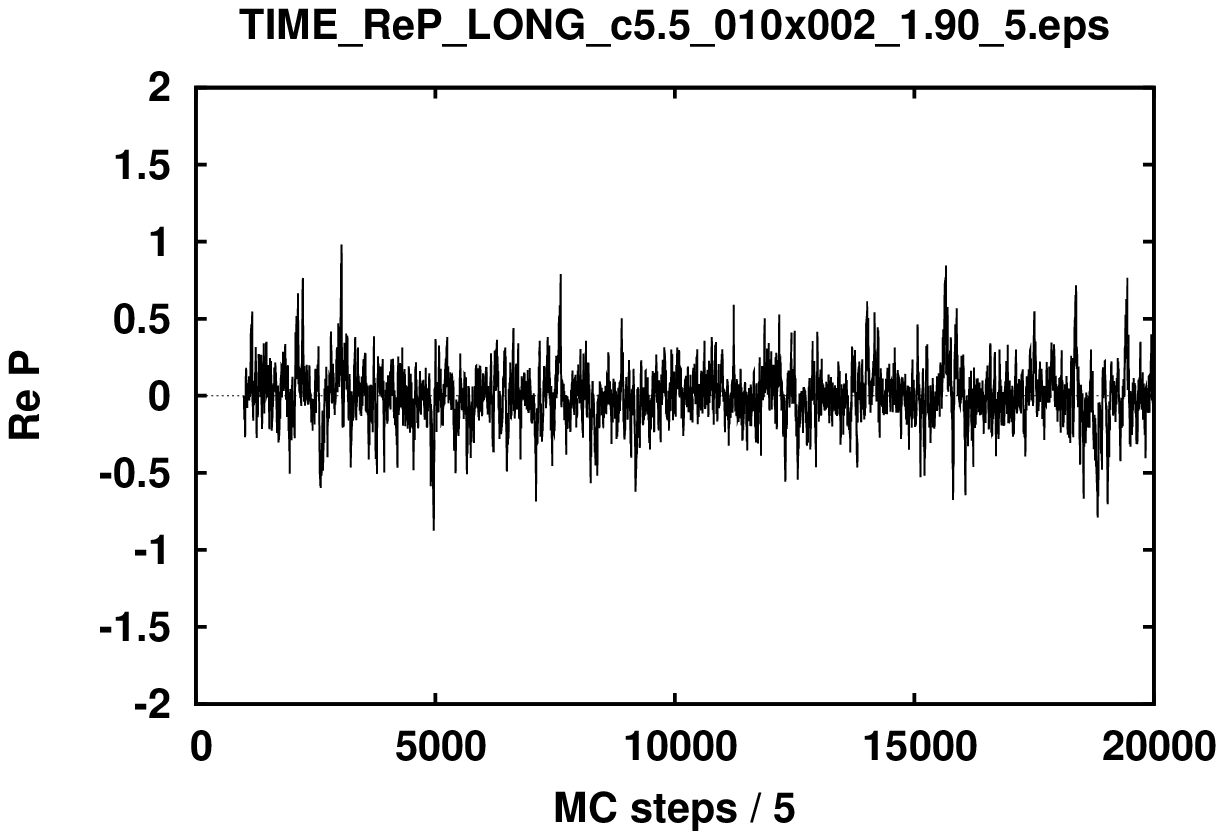} \hspace{4mm} \includegraphics[width=0.44\textwidth]{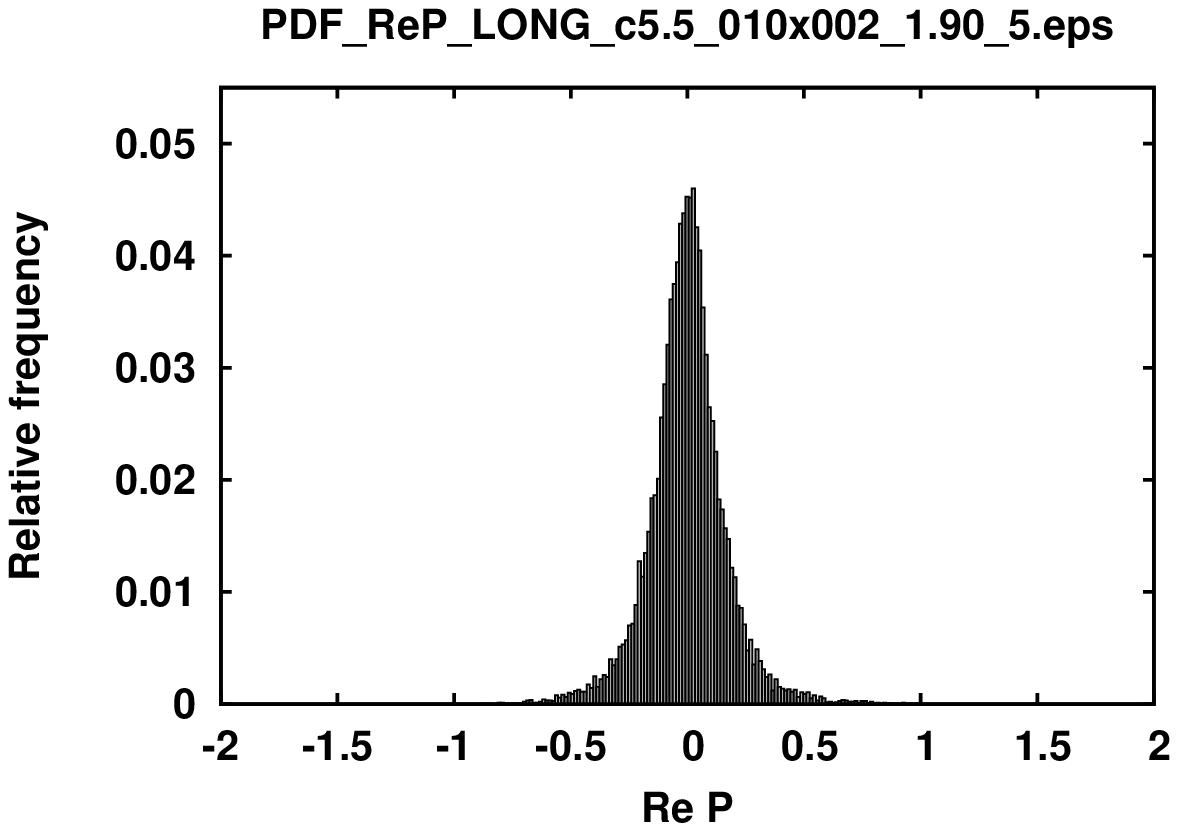}
	\includegraphics[width=0.44\textwidth]{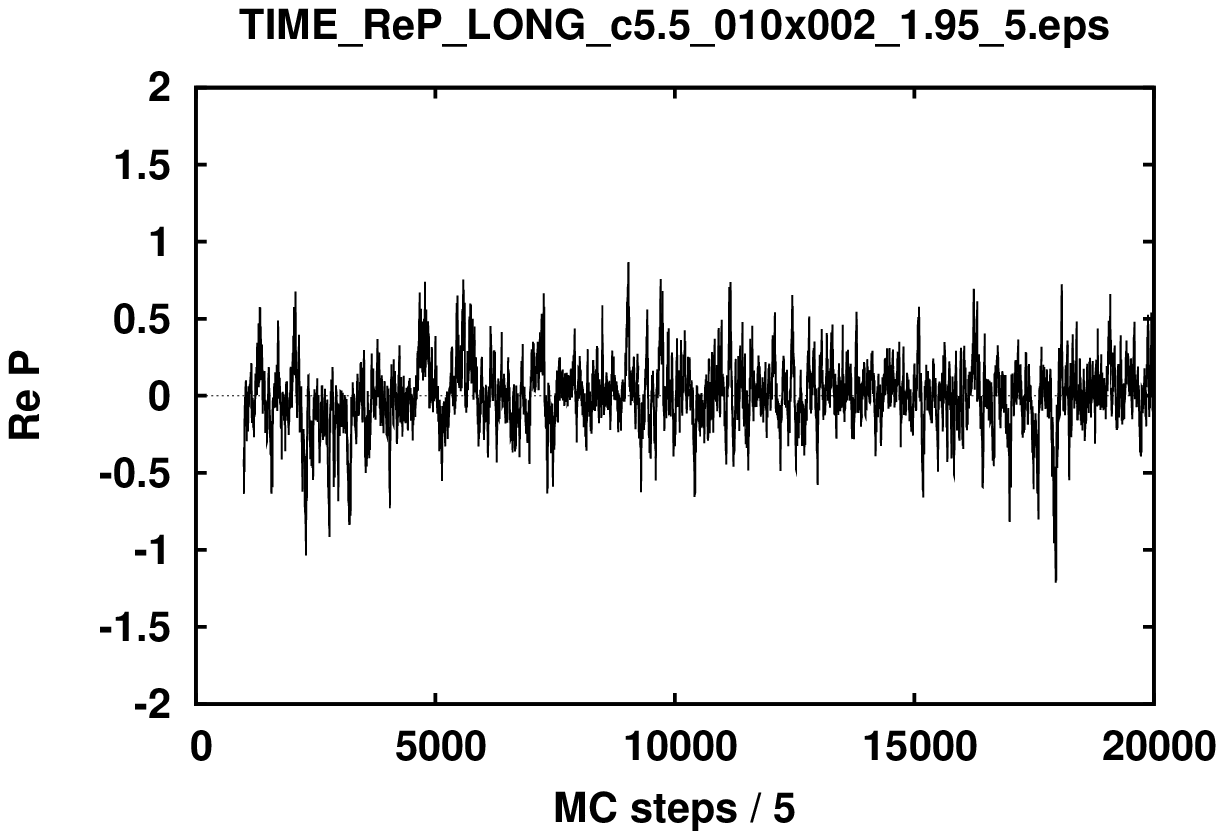} \hspace{4mm} \includegraphics[width=0.44\textwidth]{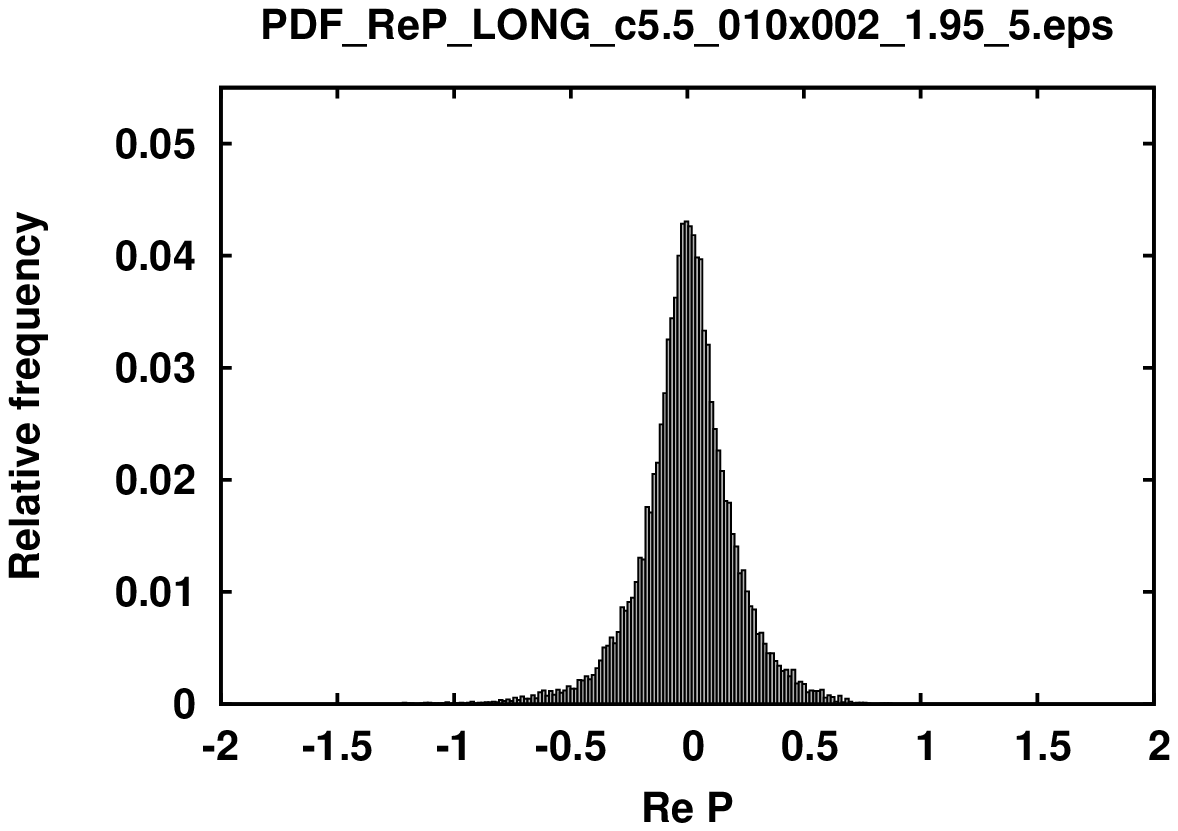}
\end{center}
\caption{ \label{FigReP_1.80_1.95}
	The Monte Carlo evolution and distribution of $\rep$ for the
	couplings $4/g^2 = 1.80, 1.85, 1.90$ and $1.95$ using $c=5.5$ from the top to the bottom. Confinement phase.
}

\end{figure}

\begin{figure}
\begin{center}
	\includegraphics[width=0.44\textwidth]{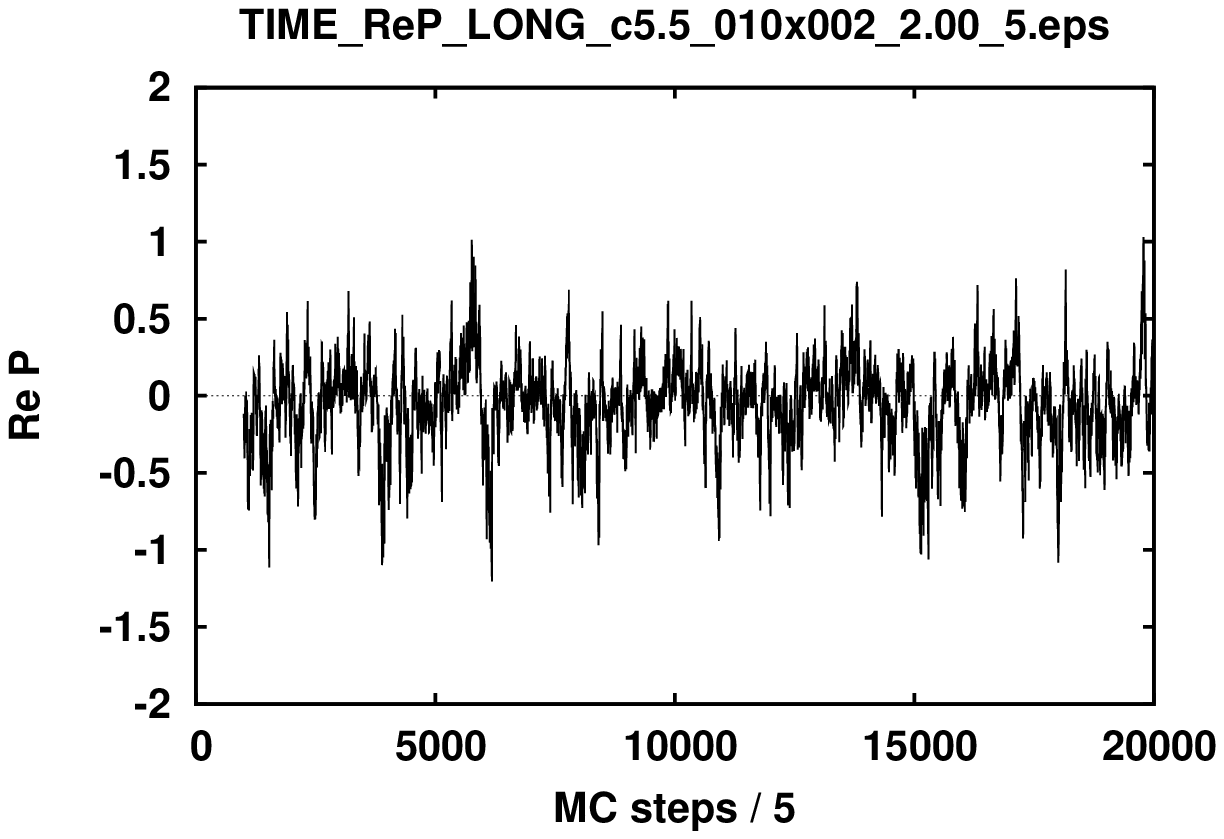} \hspace{4mm} \includegraphics[width=0.44\textwidth]{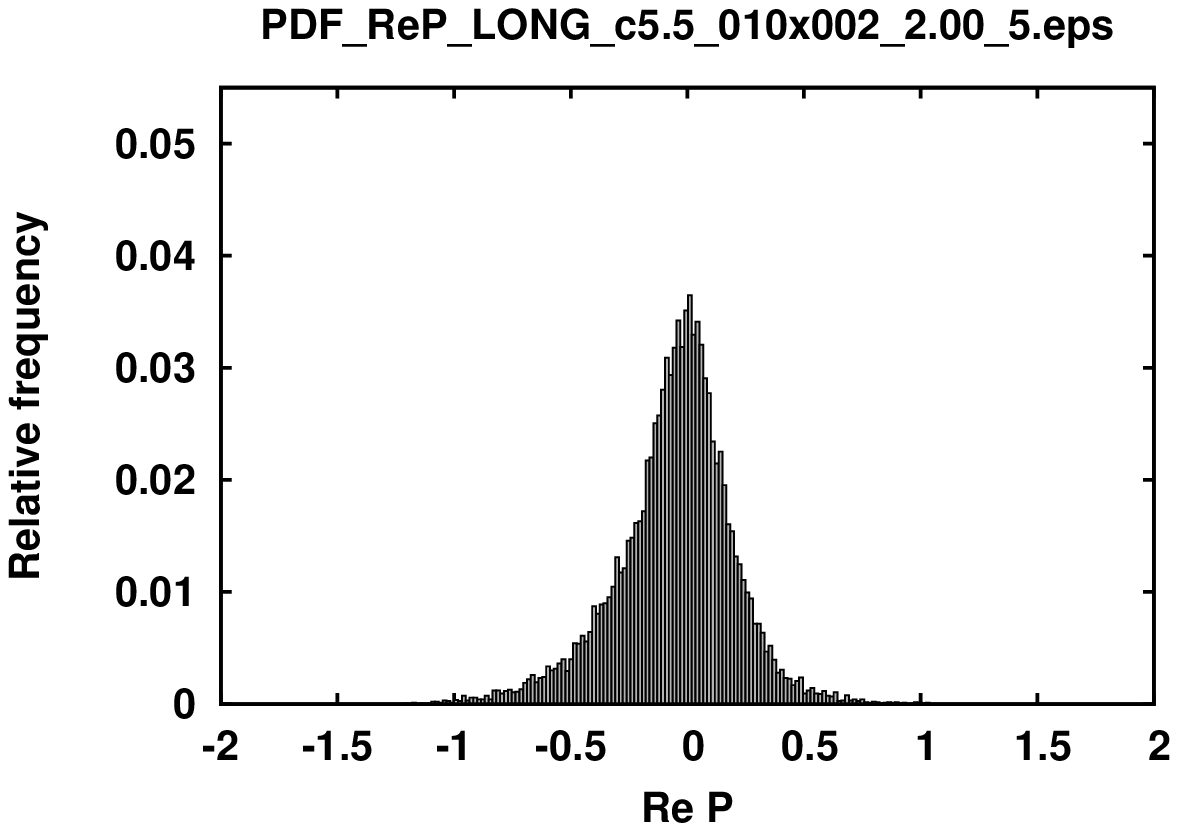}
	\includegraphics[width=0.44\textwidth]{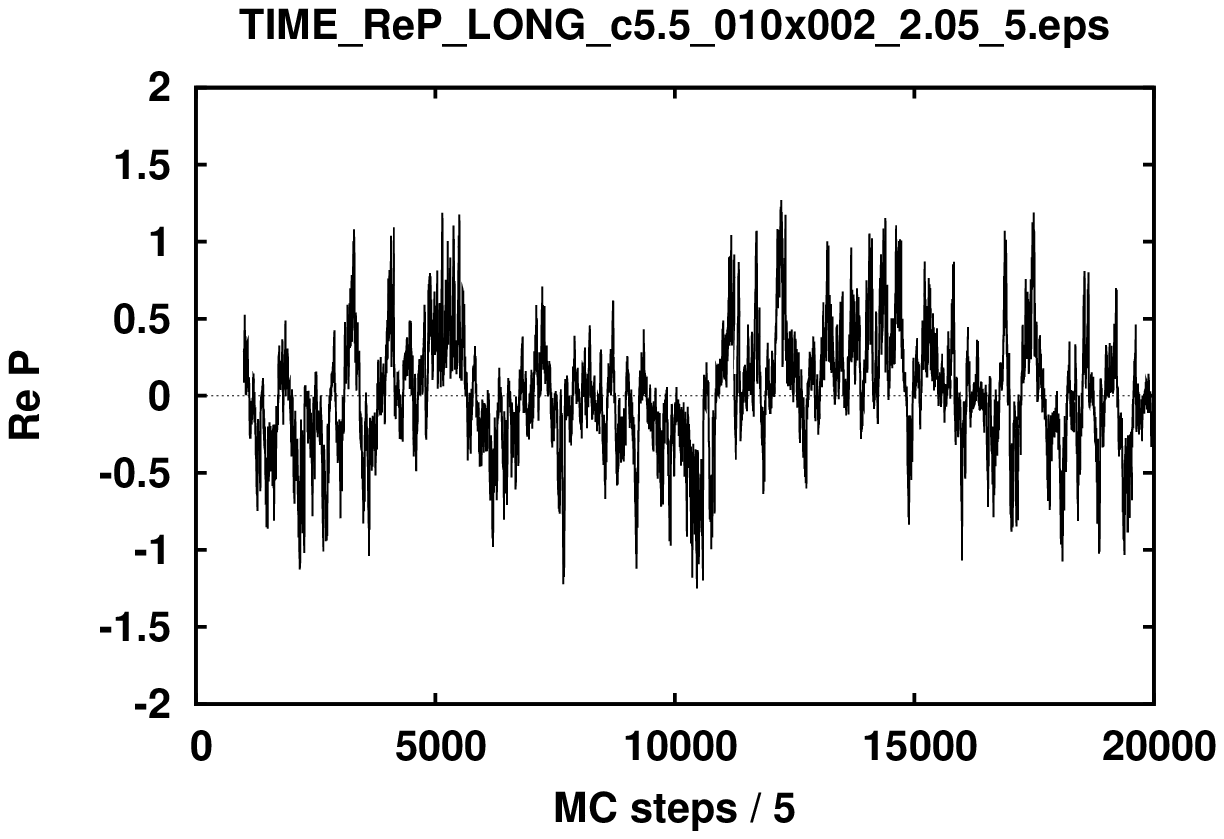} \hspace{4mm} \includegraphics[width=0.44\textwidth]{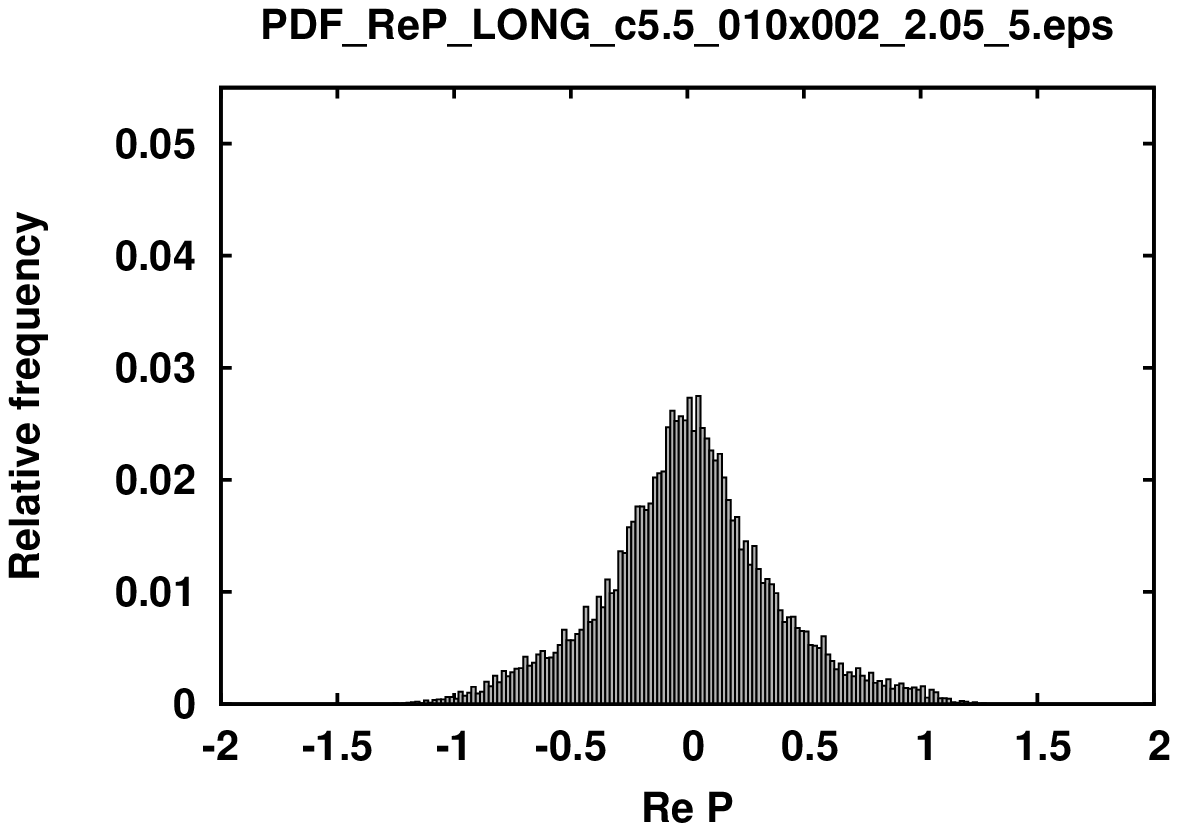}
	\includegraphics[width=0.44\textwidth]{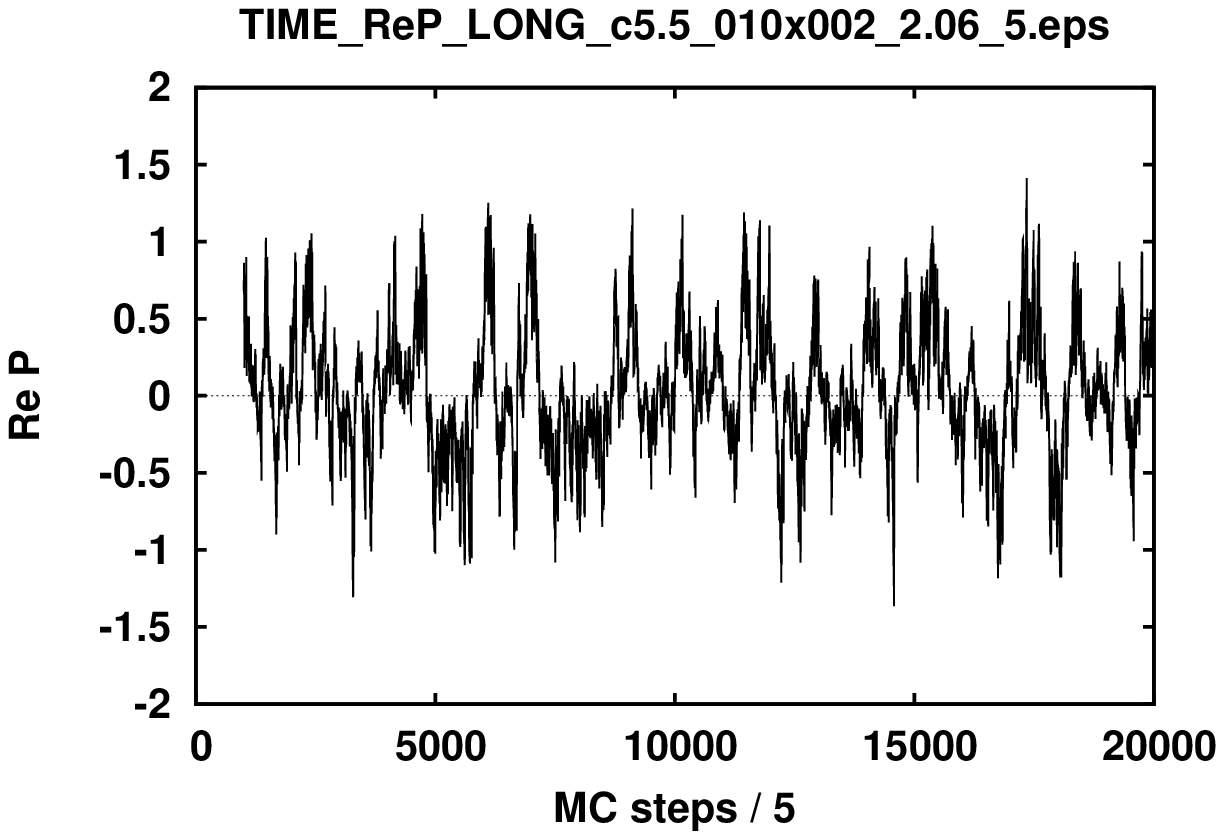} \hspace{4mm} \includegraphics[width=0.44\textwidth]{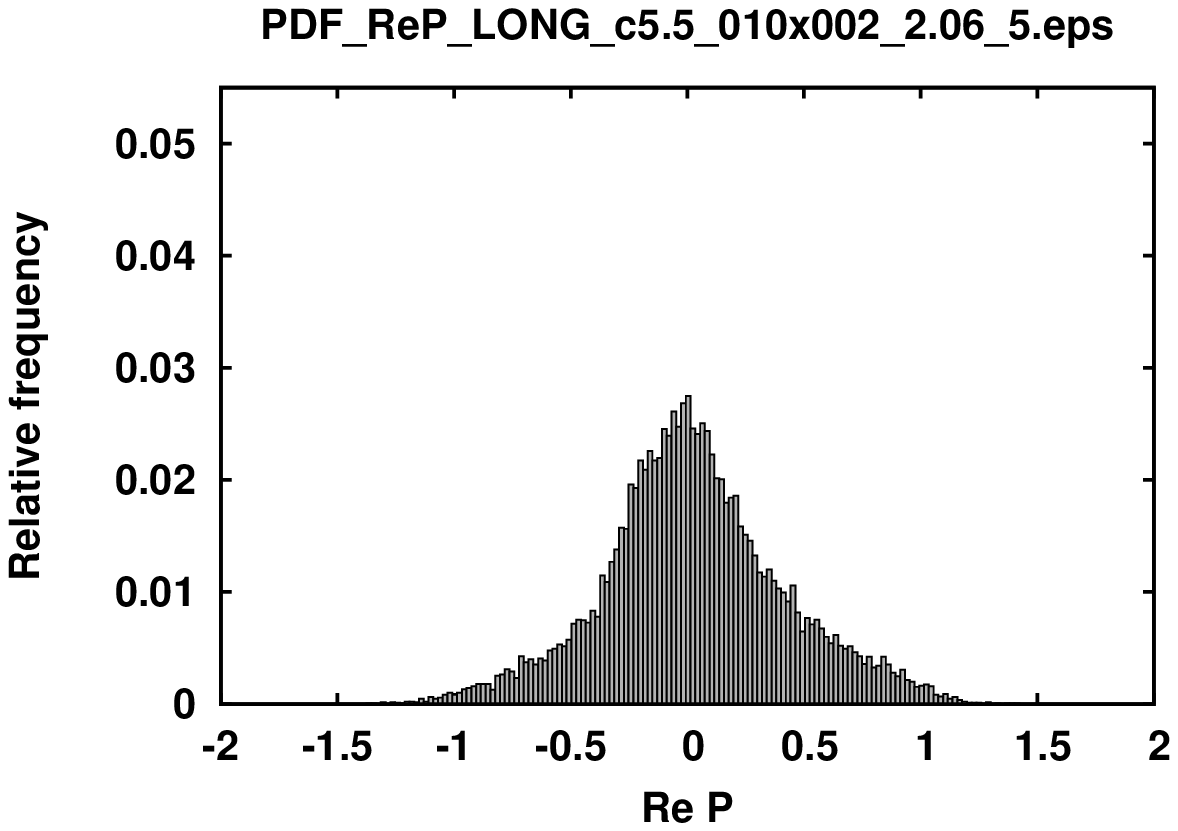}
	\includegraphics[width=0.44\textwidth]{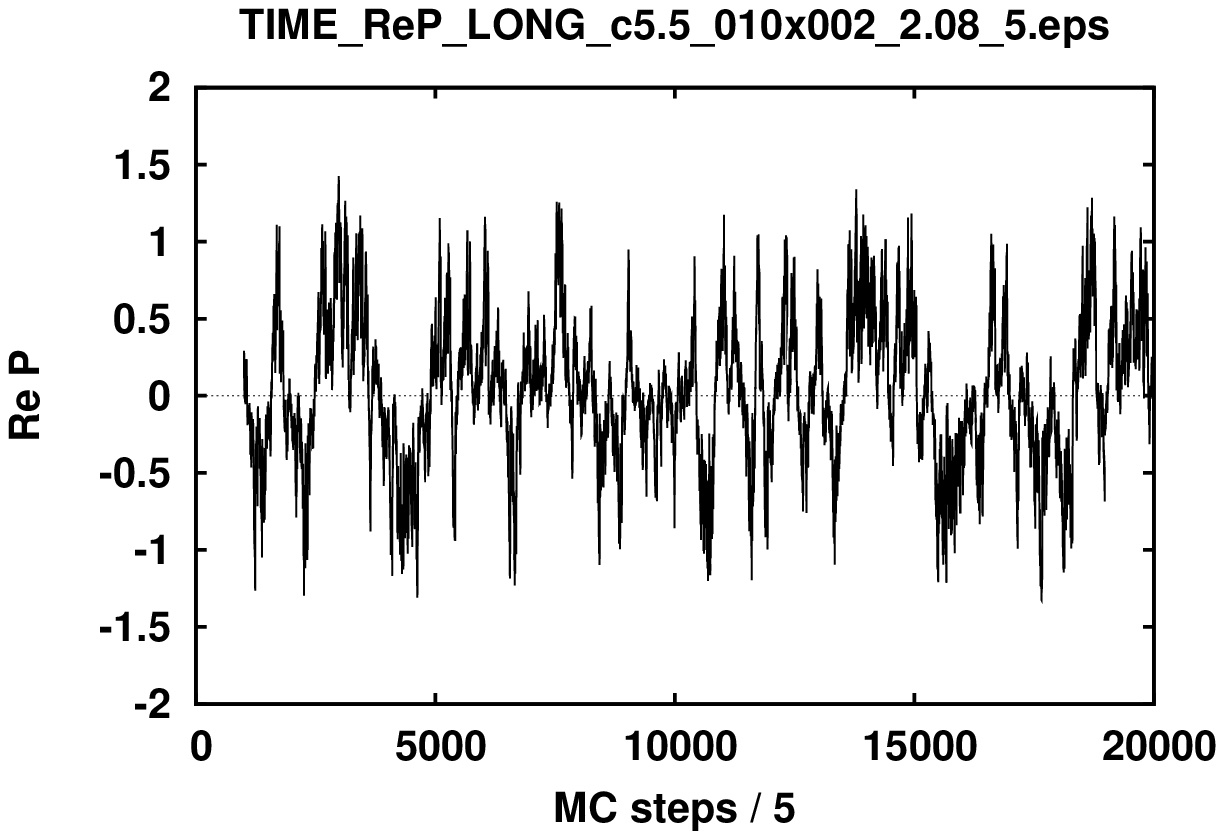} \hspace{4mm} \includegraphics[width=0.44\textwidth]{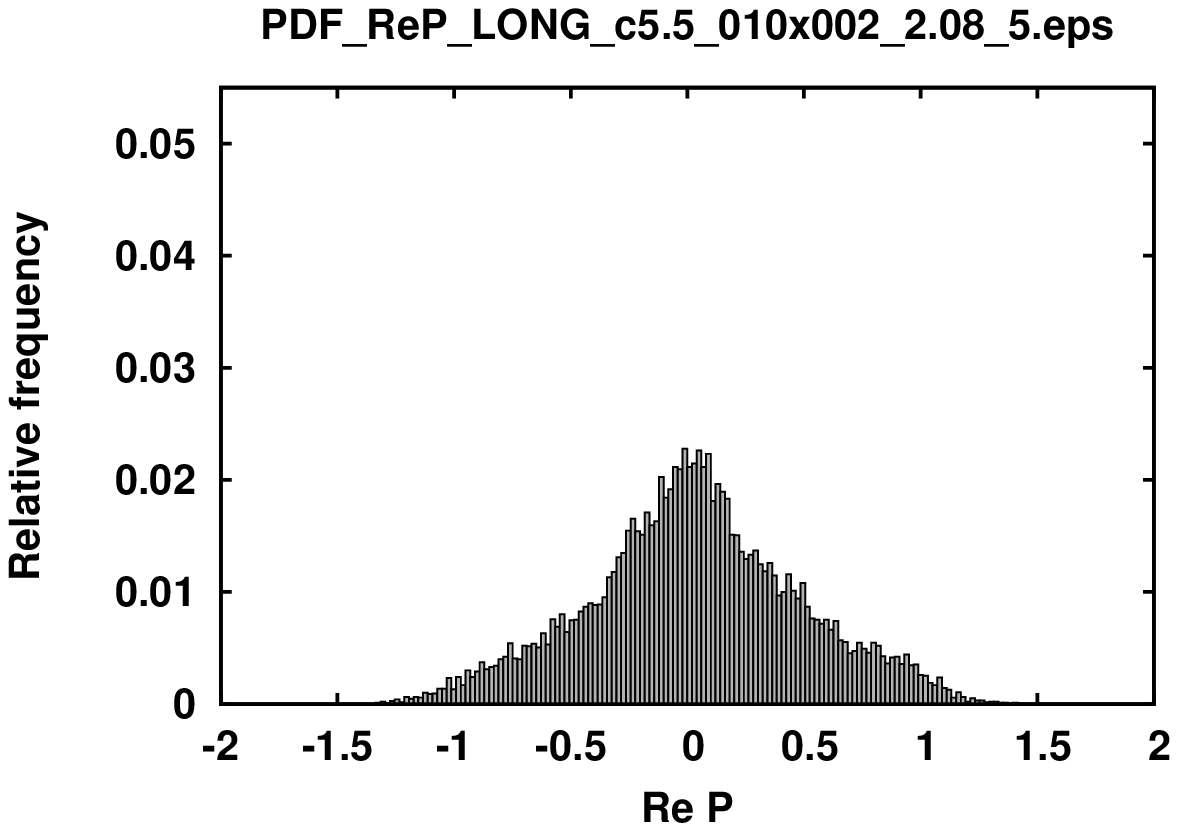}
\end{center}
\caption{ \label{FigReP_2.00_2.08}
	The Monte Carlo evolution and distribution of $\rep$ for the
	couplings $4/g^2 = 2.00, 2.05, 2.06$ and $2.08$i using $c=5.5$ from the top to the bottom.
	These couplings are nearly critical.
}

\end{figure}

\begin{figure}
\begin{center}
	\includegraphics[width=0.44\textwidth]{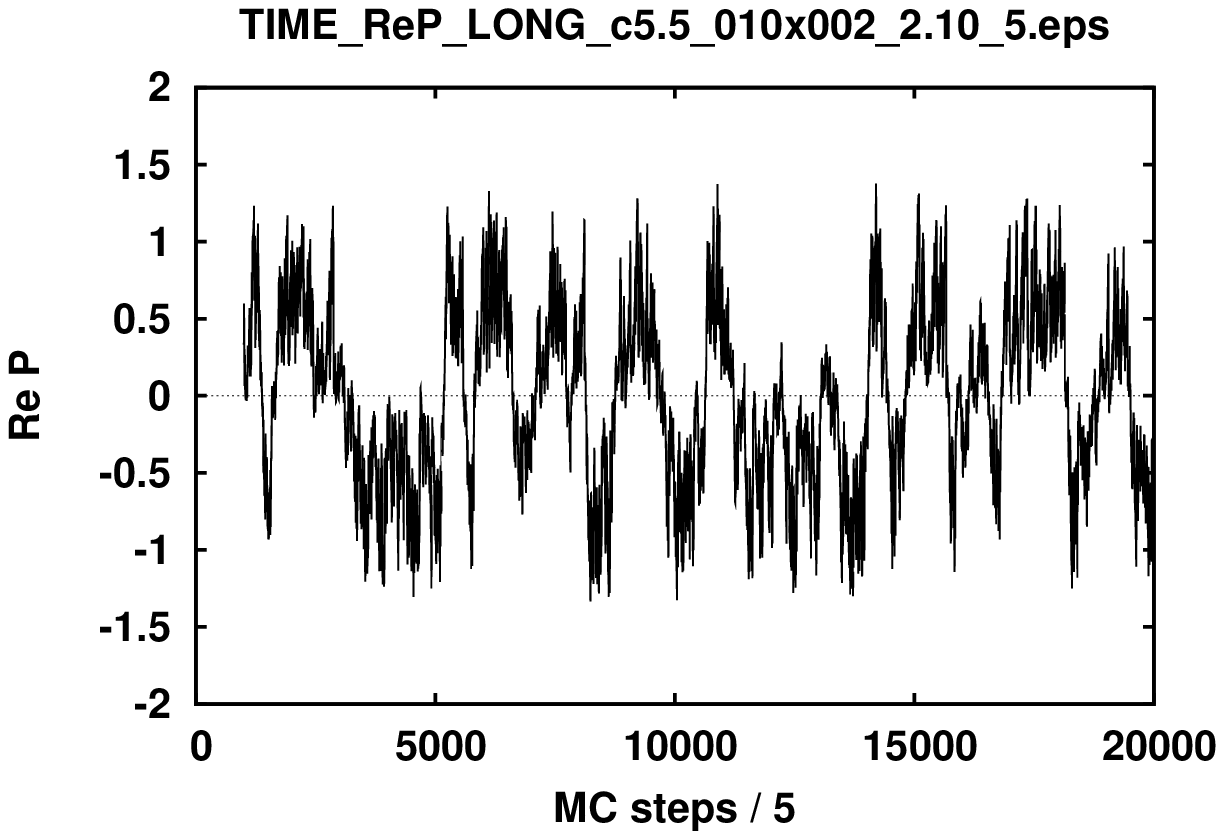} \hspace{4mm} \includegraphics[width=0.44\textwidth]{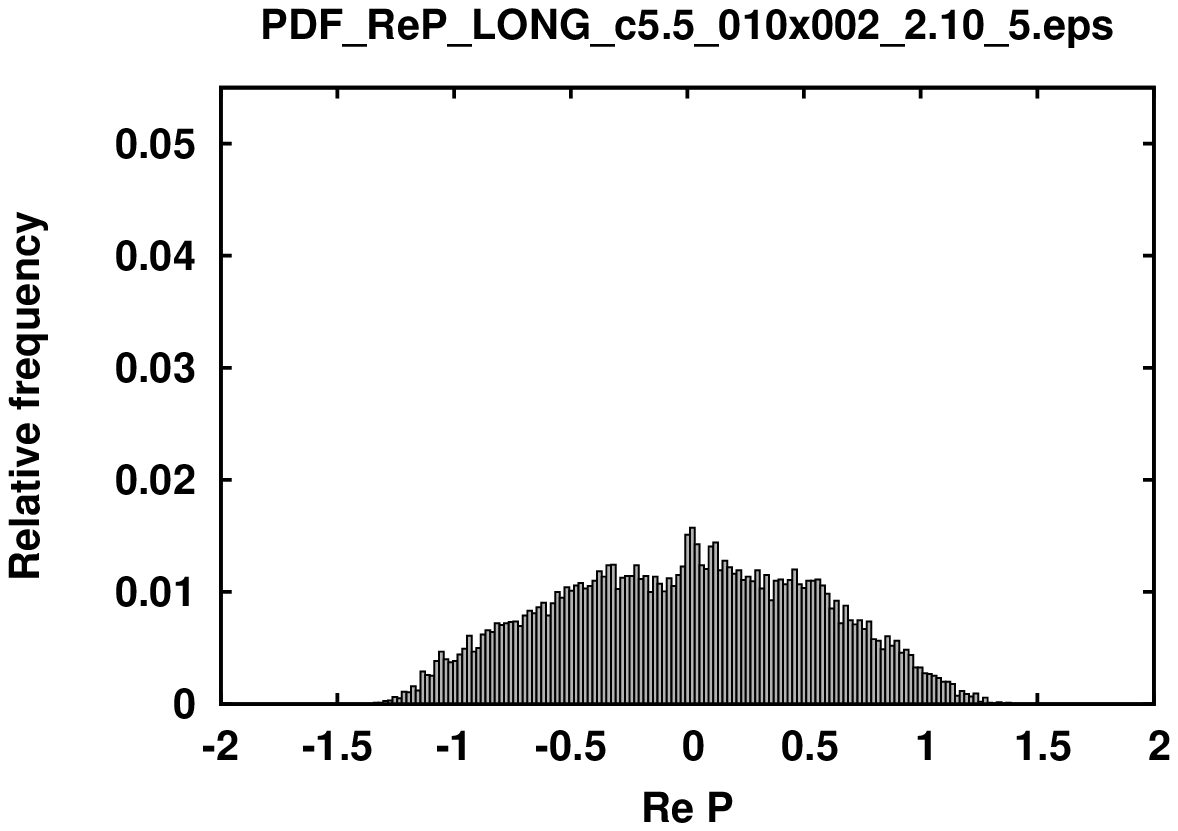}
	\includegraphics[width=0.44\textwidth]{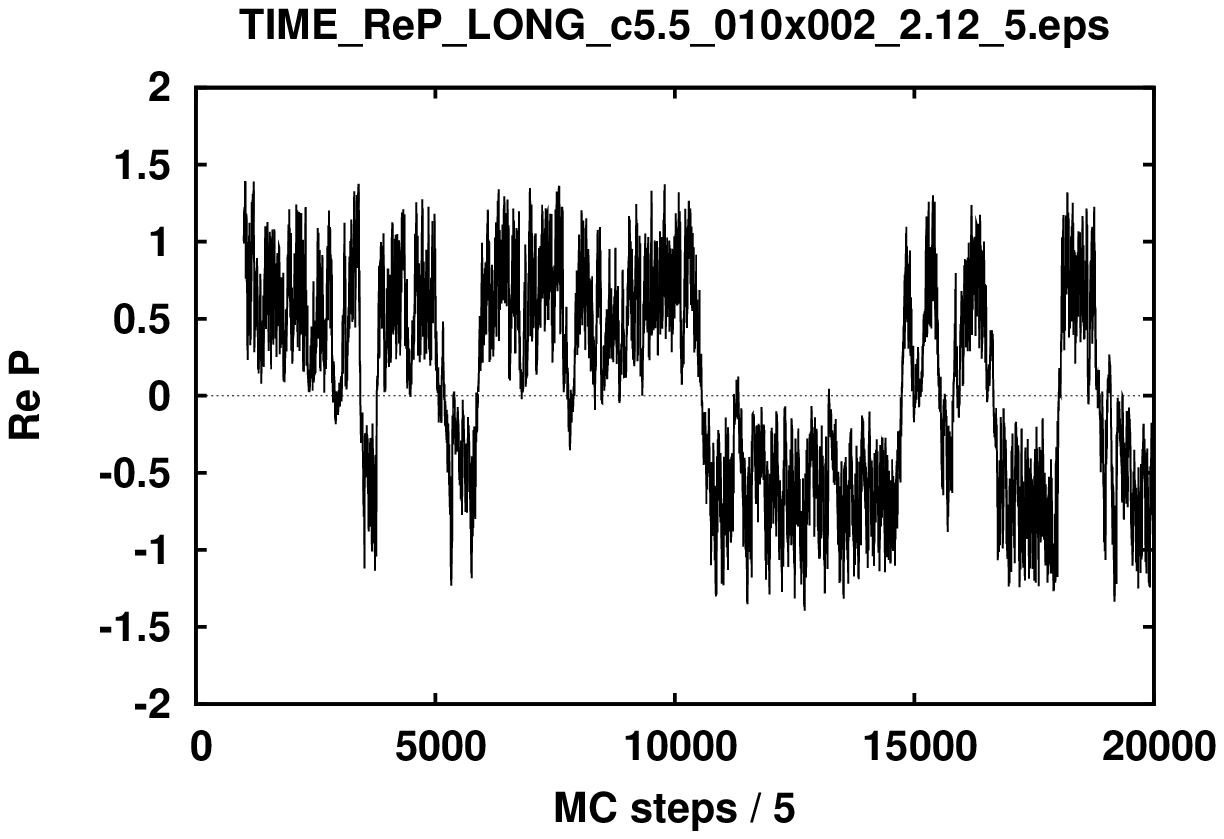} \hspace{4mm} \includegraphics[width=0.44\textwidth]{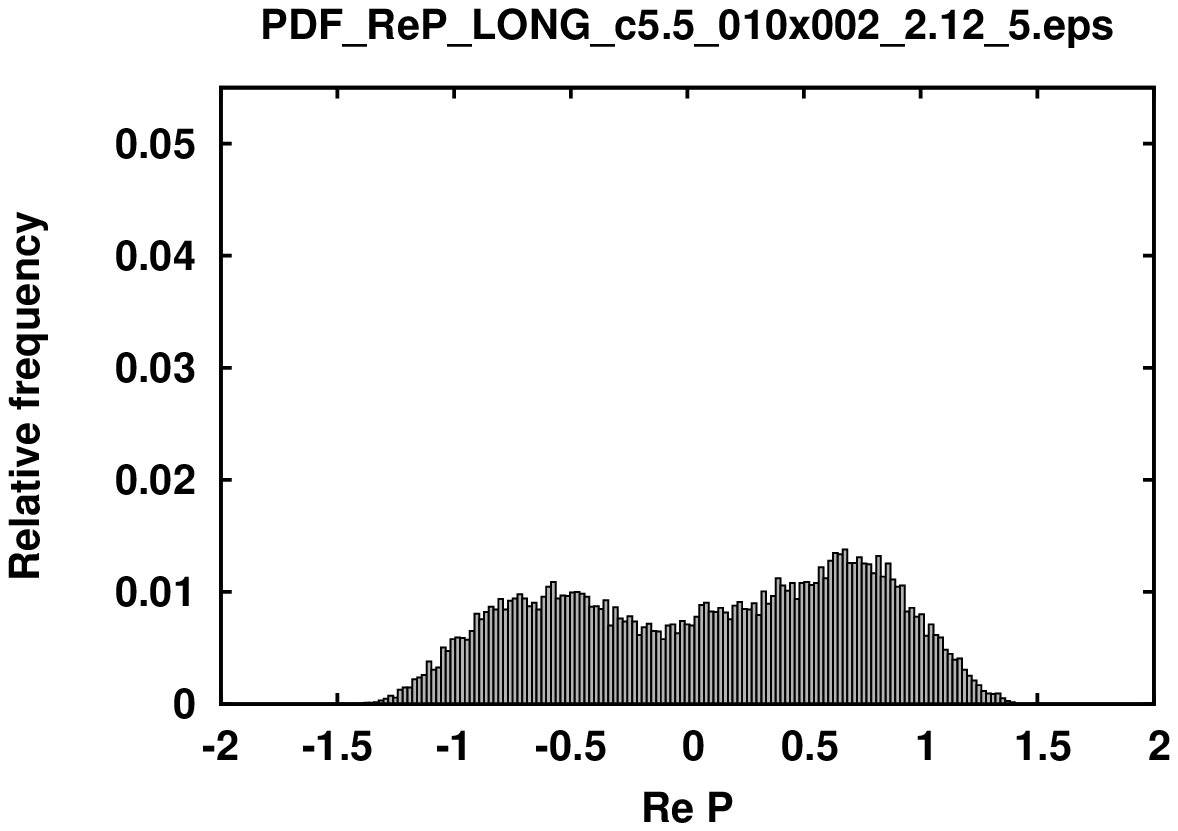}
	\includegraphics[width=0.44\textwidth]{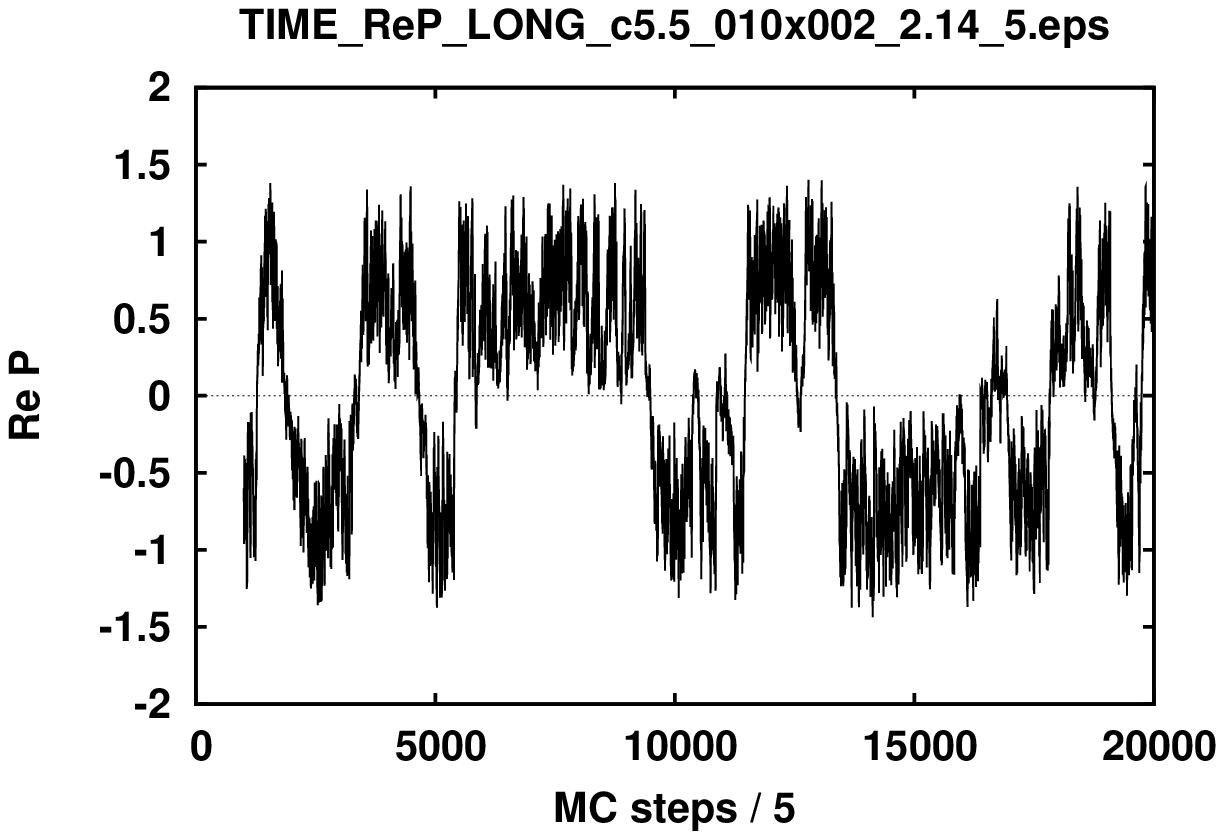} \hspace{4mm} \includegraphics[width=0.44\textwidth]{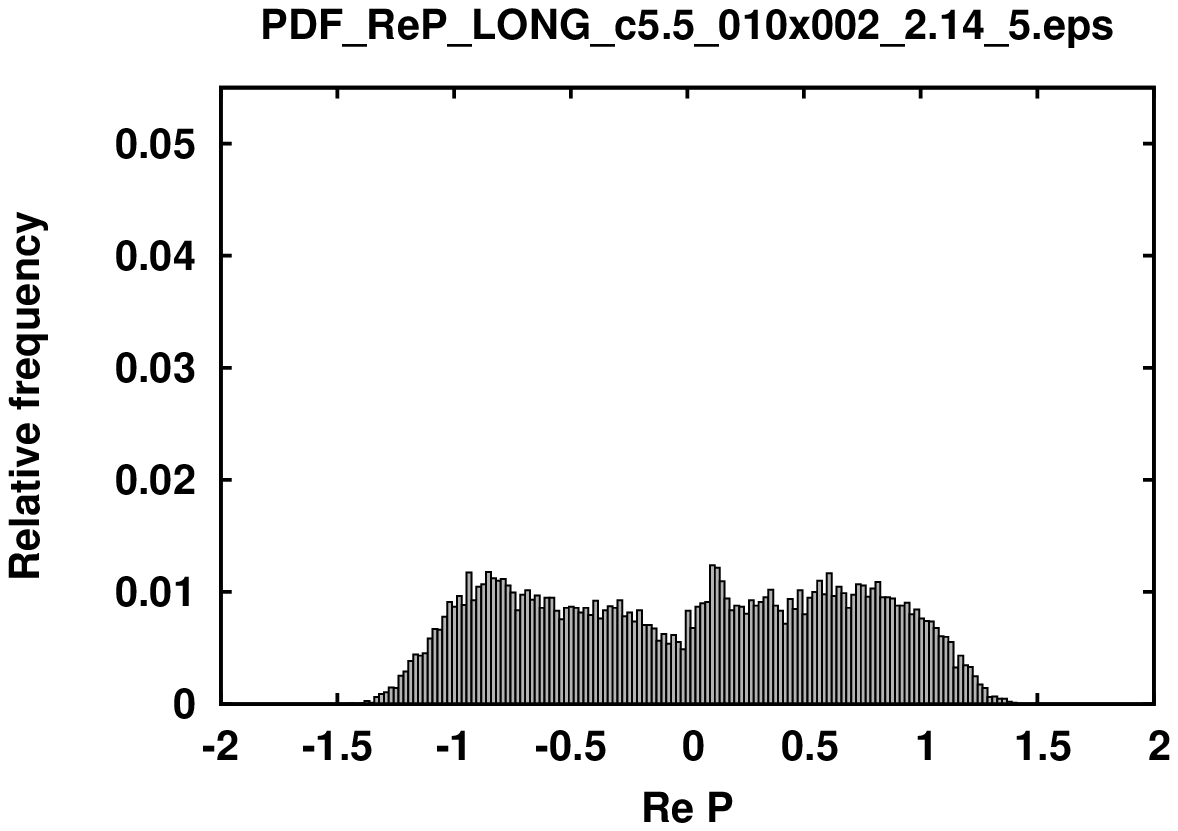}
	\includegraphics[width=0.44\textwidth]{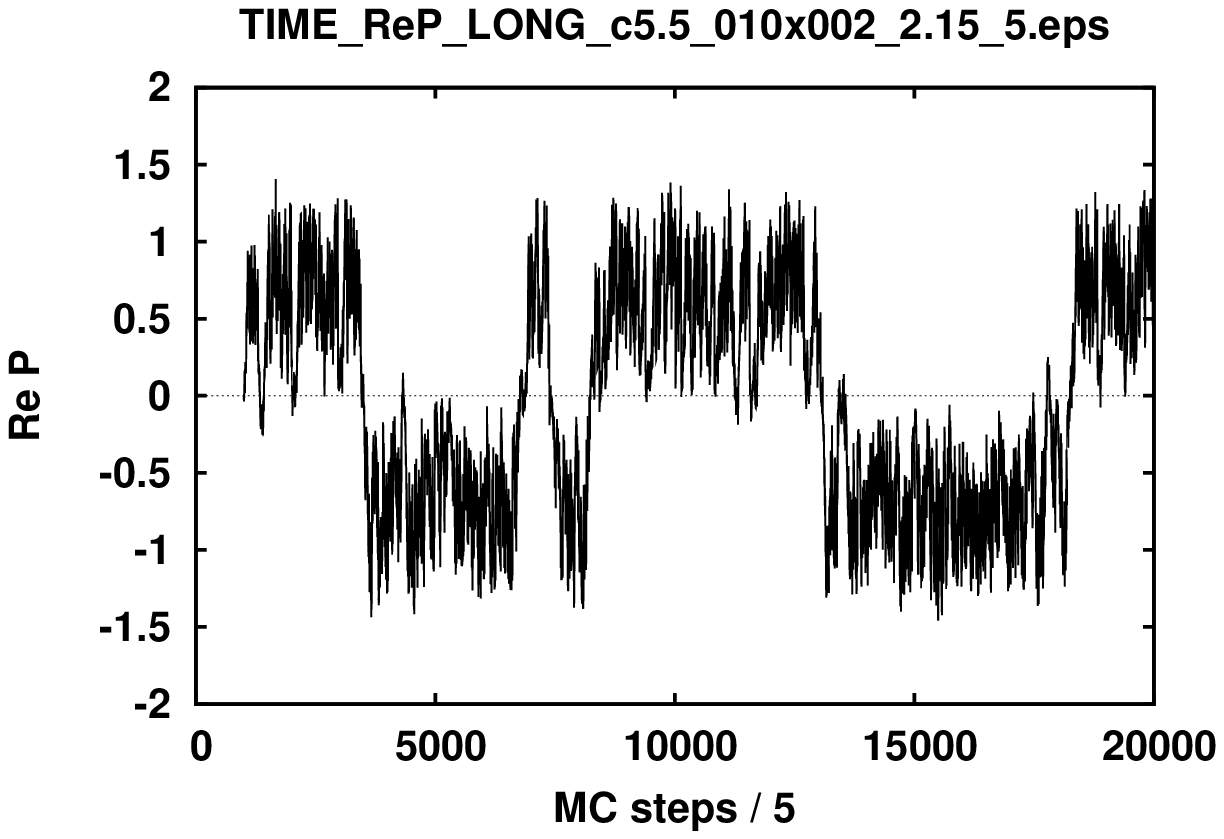} \hspace{4mm} \includegraphics[width=0.44\textwidth]{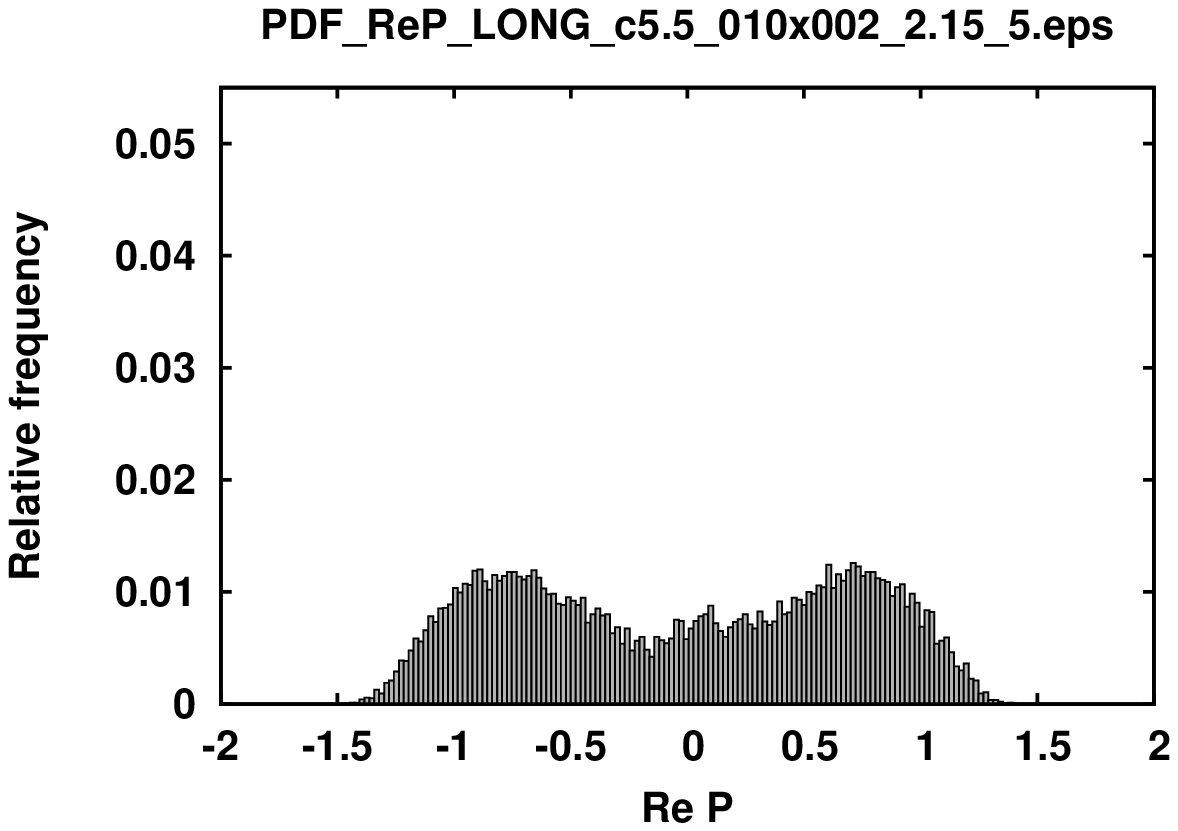}
\end{center}
\caption{ \label{FigReP_2.10_2.15}
	The Monte Carlo evolution and distribution of $\rep$ for the
	couplings $4/g^2 = 2.10, 2.12, 2.14$ and $2.15$ using $c=5.5$ from the top to the bottom.
	Here the two-peak distribution develops, the deconfinement sets in.
}

\end{figure}

\begin{figure}
\begin{center}
	\includegraphics[width=0.44\textwidth]{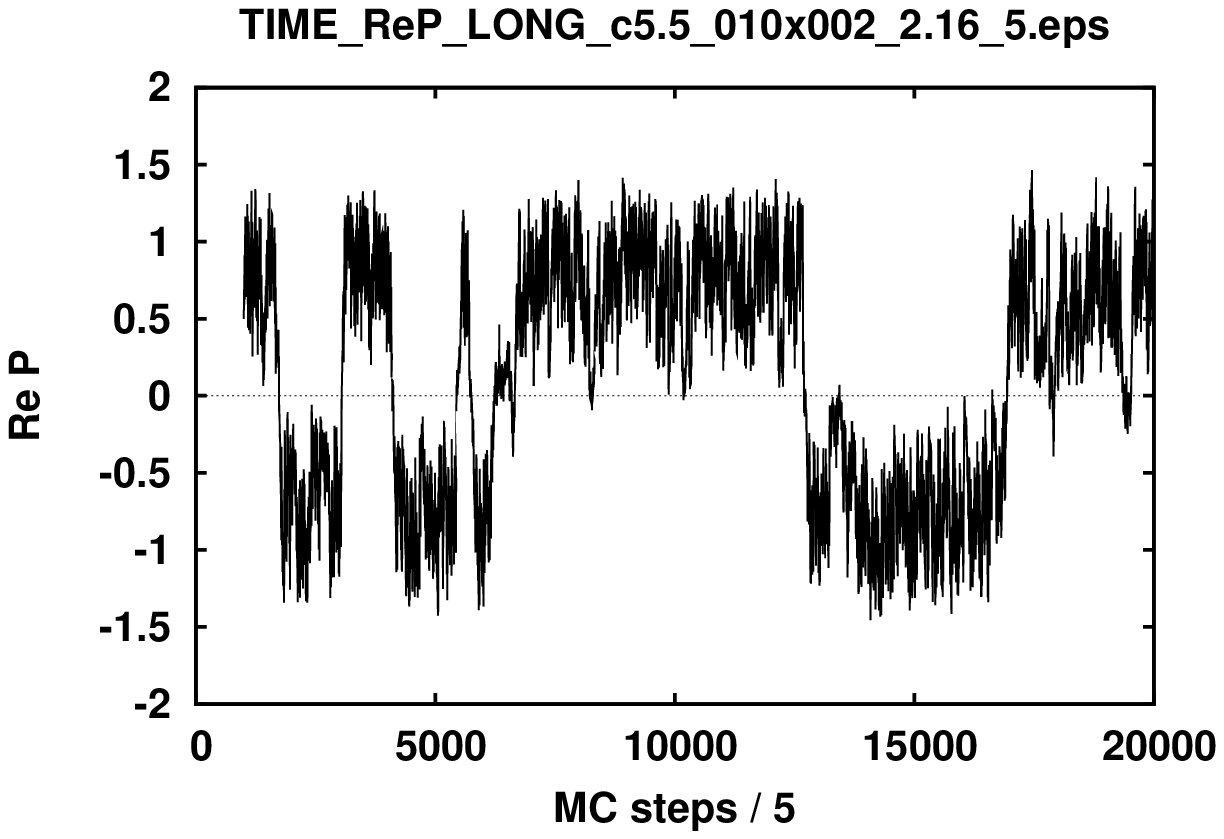} \hspace{4mm} \includegraphics[width=0.44\textwidth]{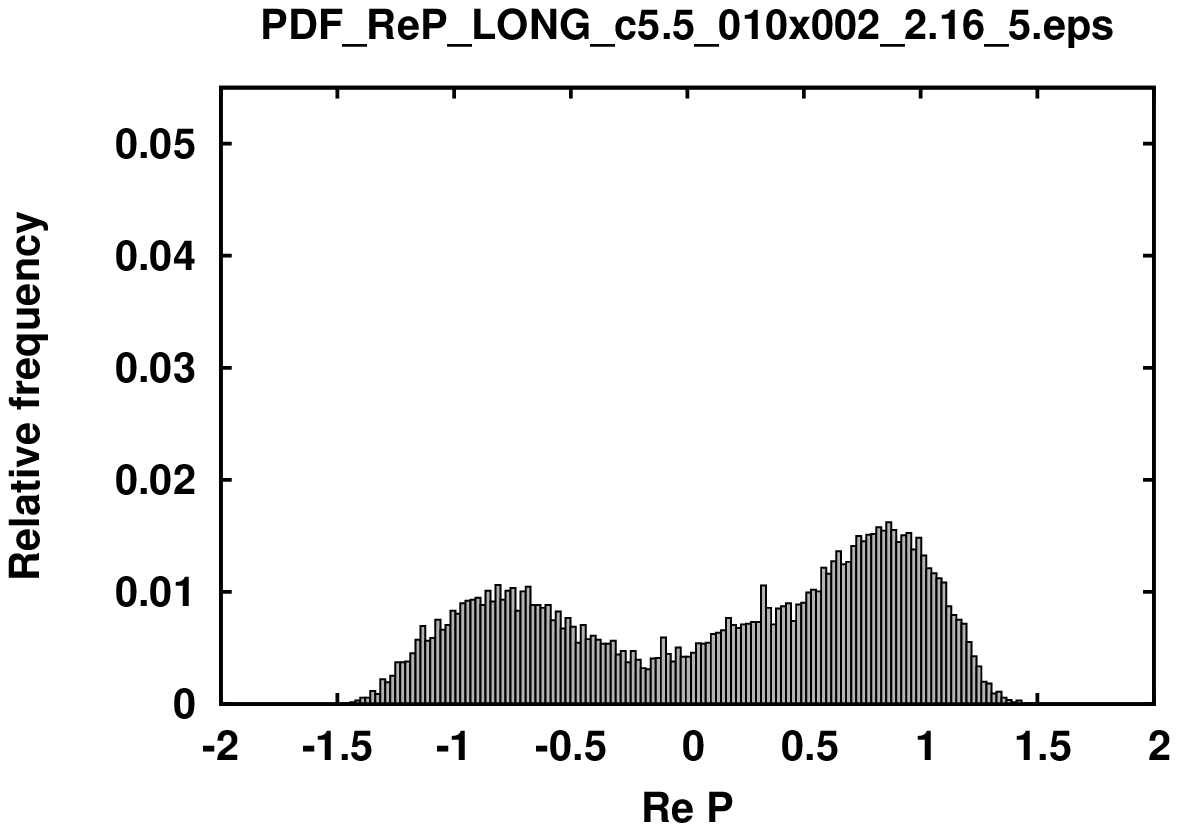}
	\includegraphics[width=0.44\textwidth]{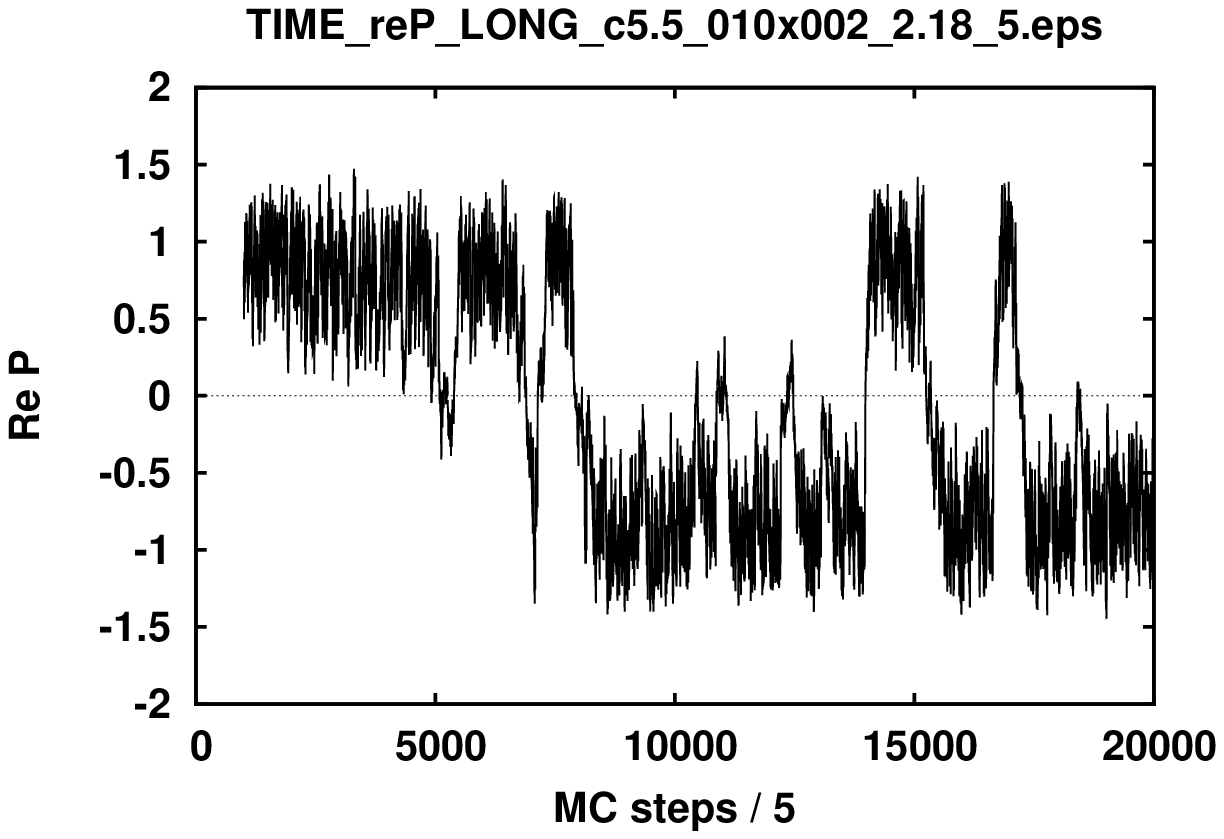} \hspace{4mm} \includegraphics[width=0.44\textwidth]{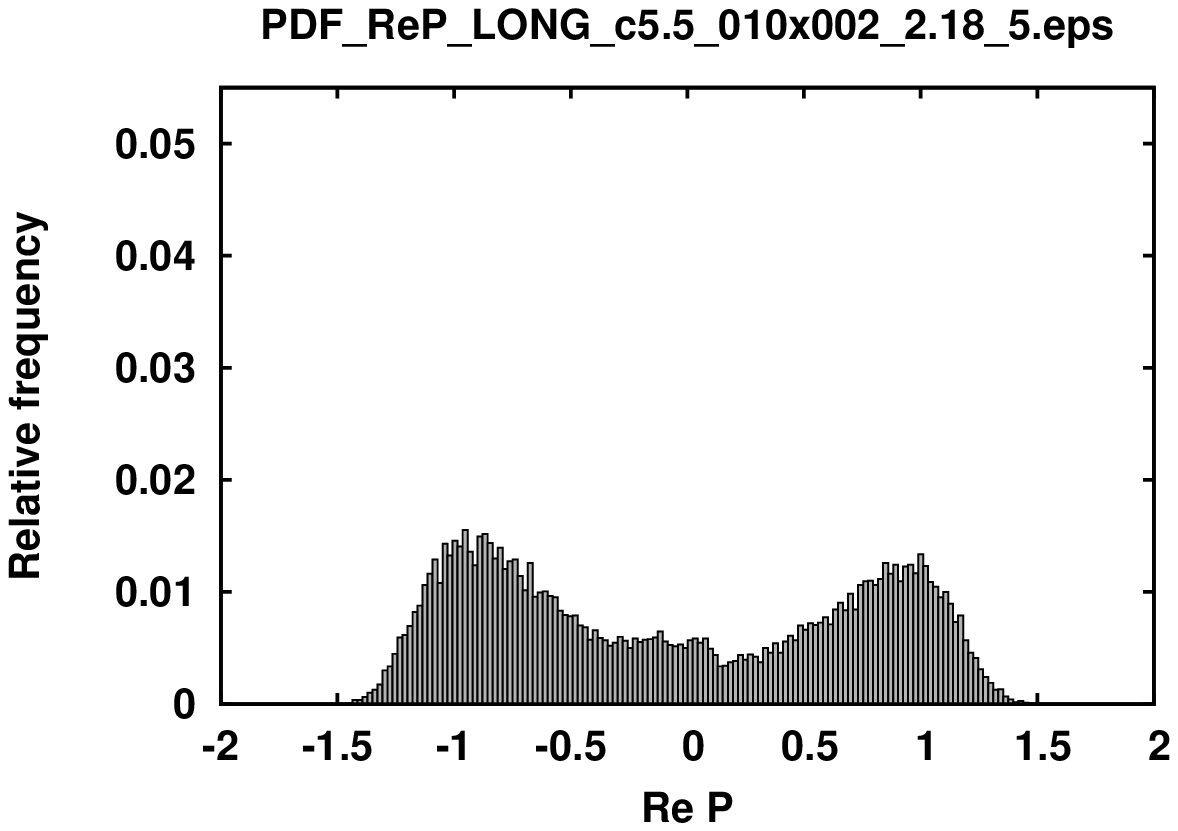}
	\includegraphics[width=0.44\textwidth]{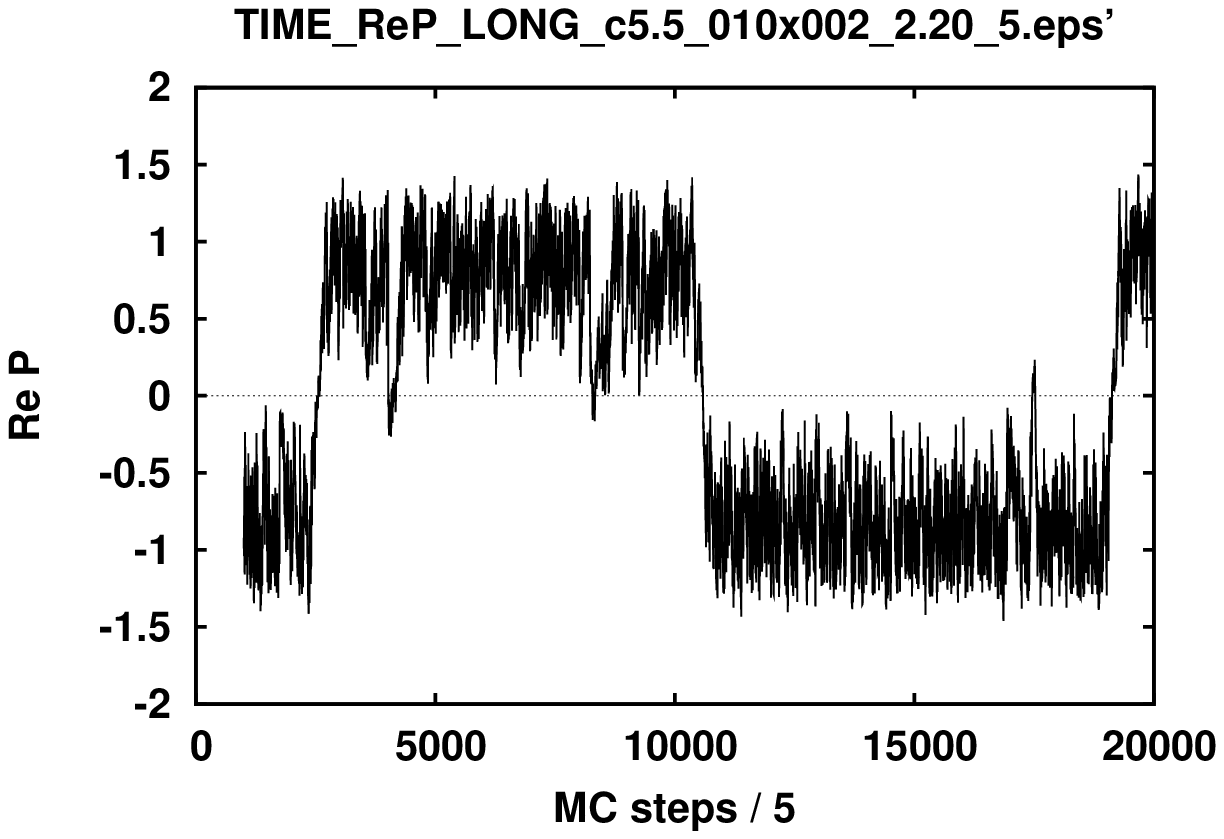} \hspace{4mm} \includegraphics[width=0.44\textwidth]{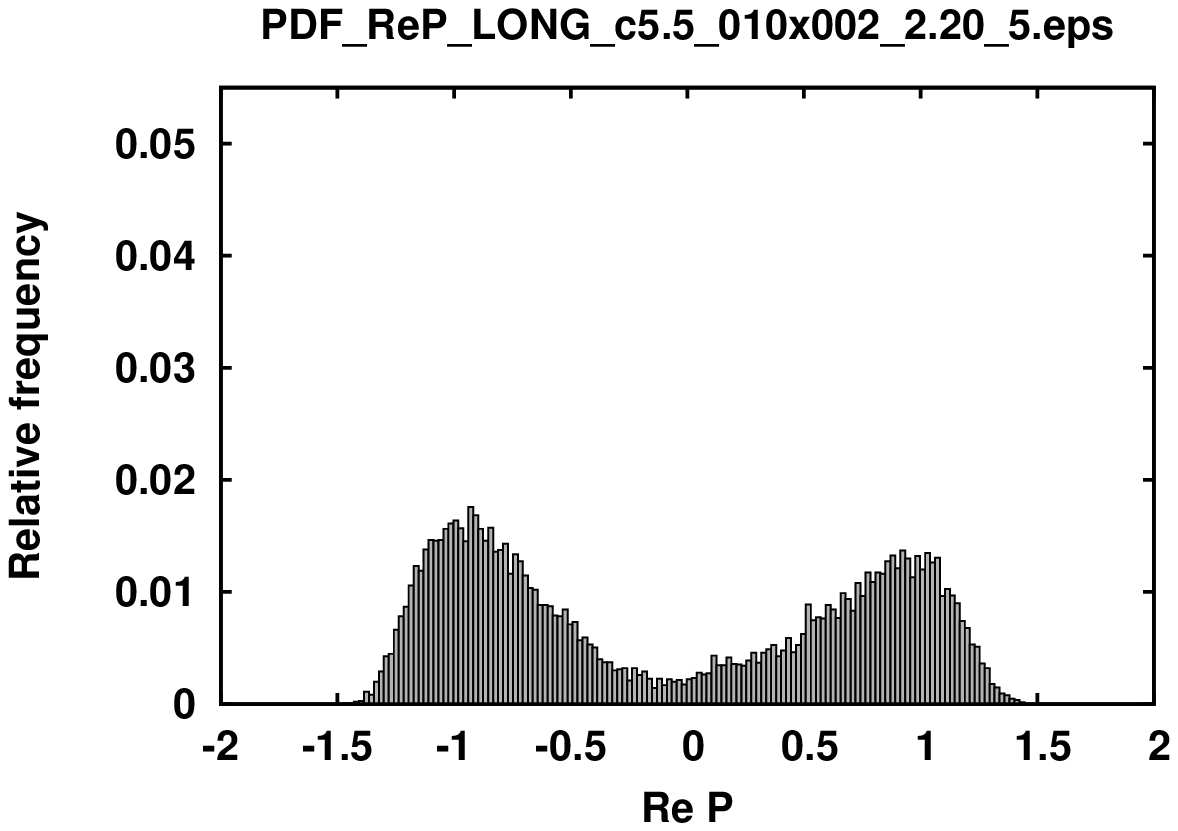}
	\includegraphics[width=0.44\textwidth]{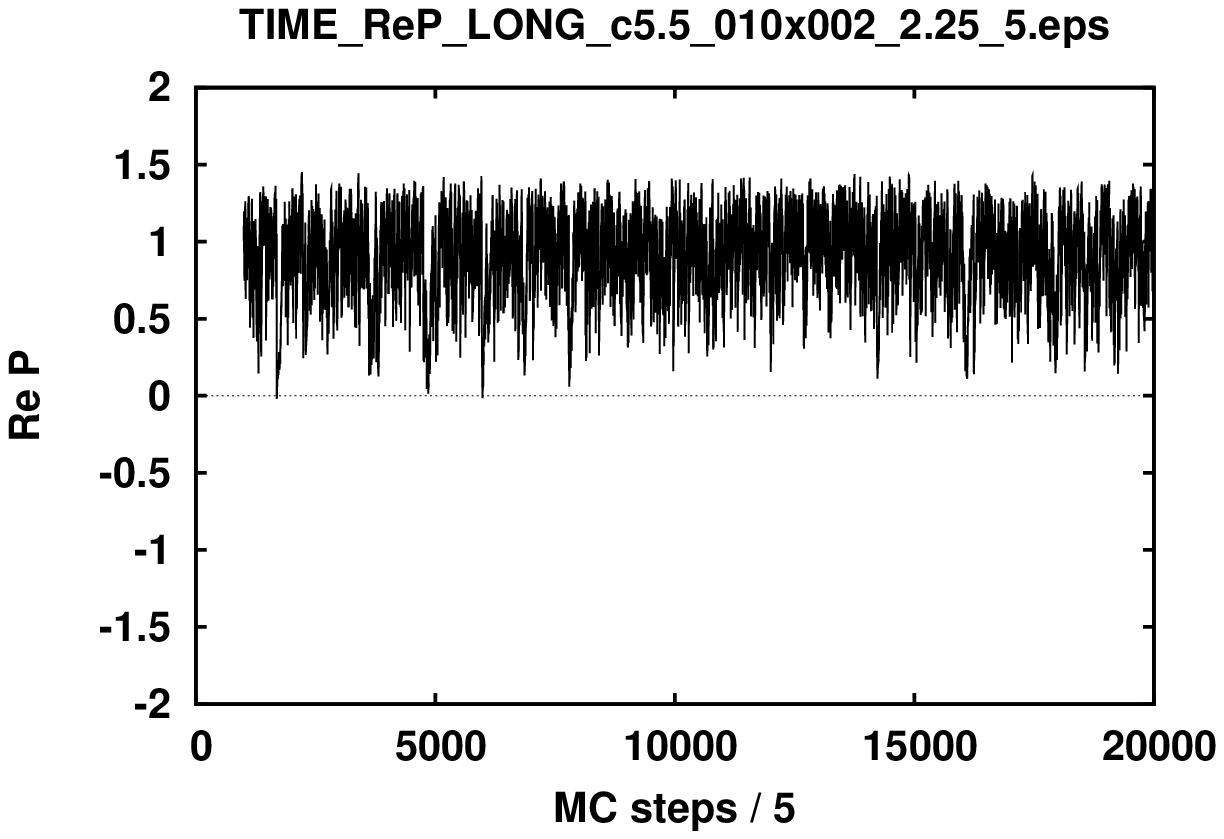} \hspace{4mm} \includegraphics[width=0.44\textwidth]{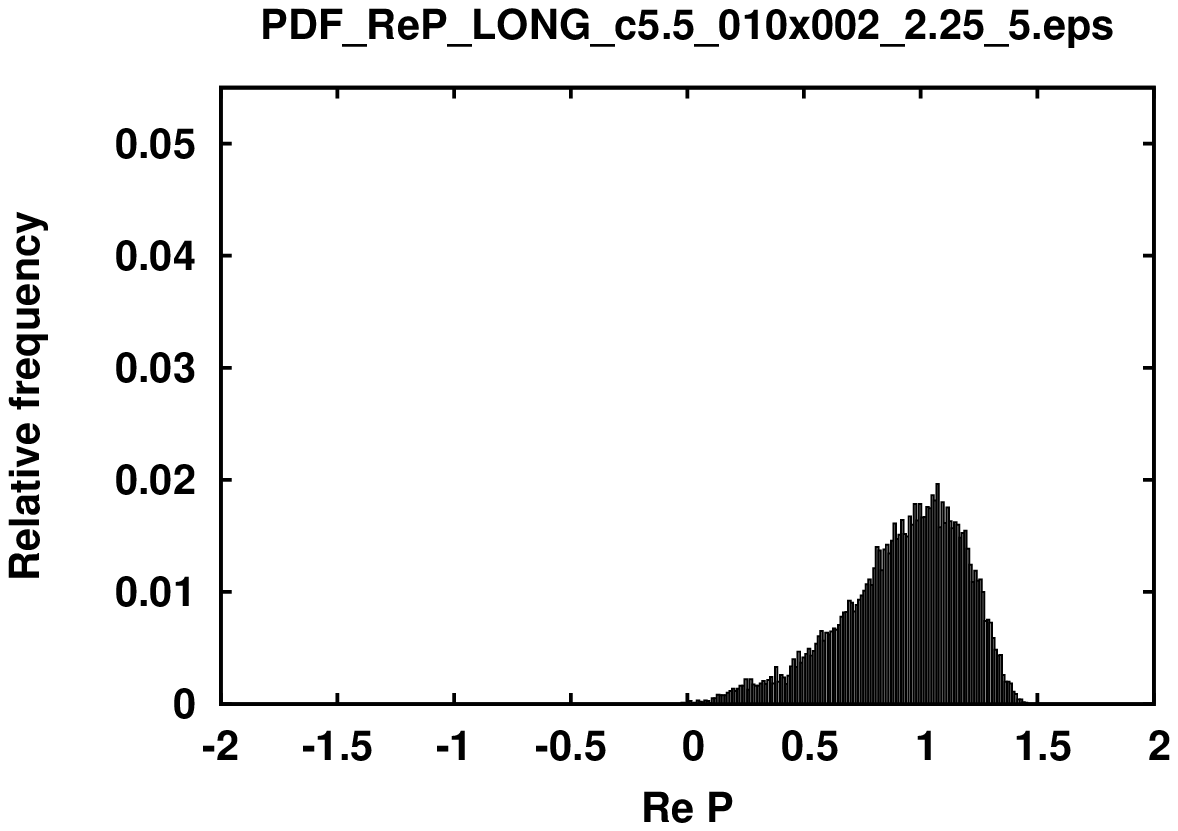}
\end{center}
\caption{ \label{FigReP_2.16_2.25}
	The Monte Carlo evolution and distribution of $\rep$ for the
	couplings $4/g^2 = 2.16, 2.18, 2.20$ and $2.25$ using $c=5.5$ from the top to the bottom.
	By these couplings we dwell into the deconfinement regime.
}

\end{figure}

\begin{figure}
\begin{center}
	\includegraphics[width=0.44\textwidth]{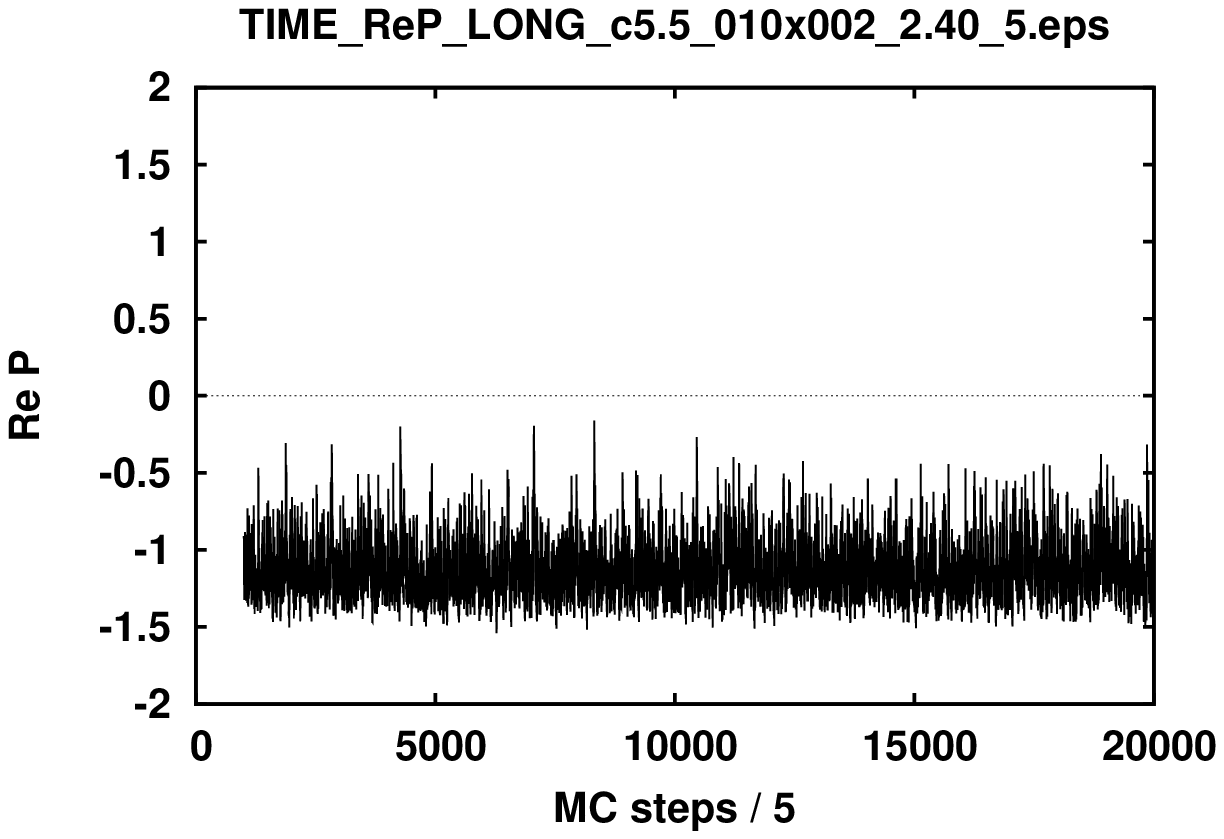} \hspace{4mm} \includegraphics[width=0.44\textwidth]{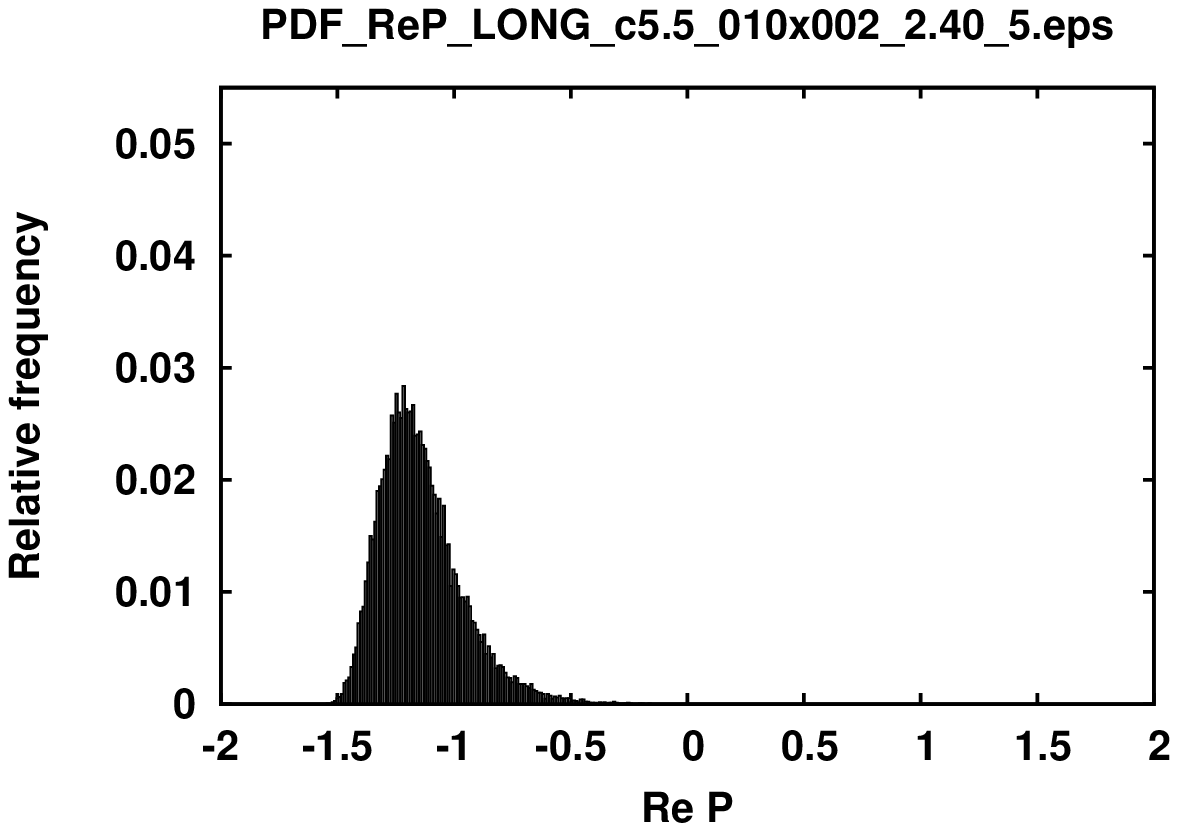}
	\includegraphics[width=0.44\textwidth]{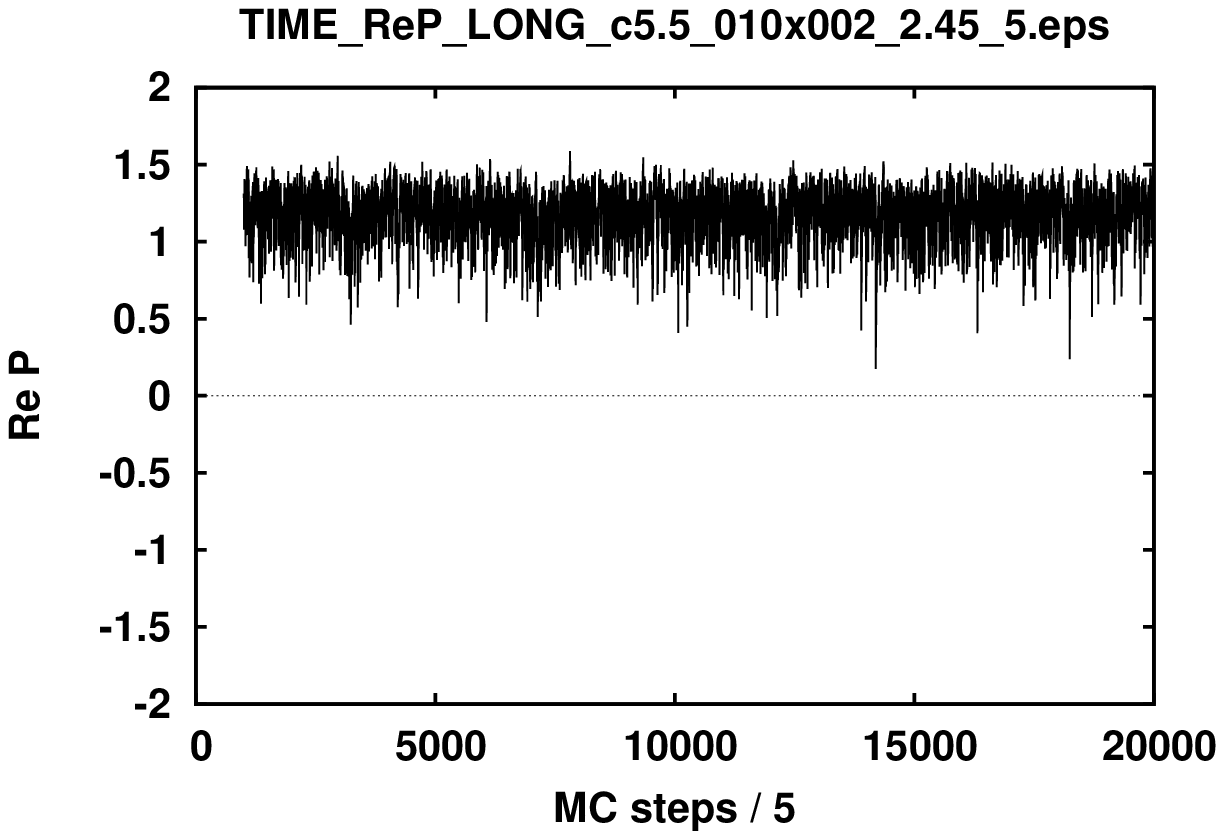} \hspace{4mm} \includegraphics[width=0.44\textwidth]{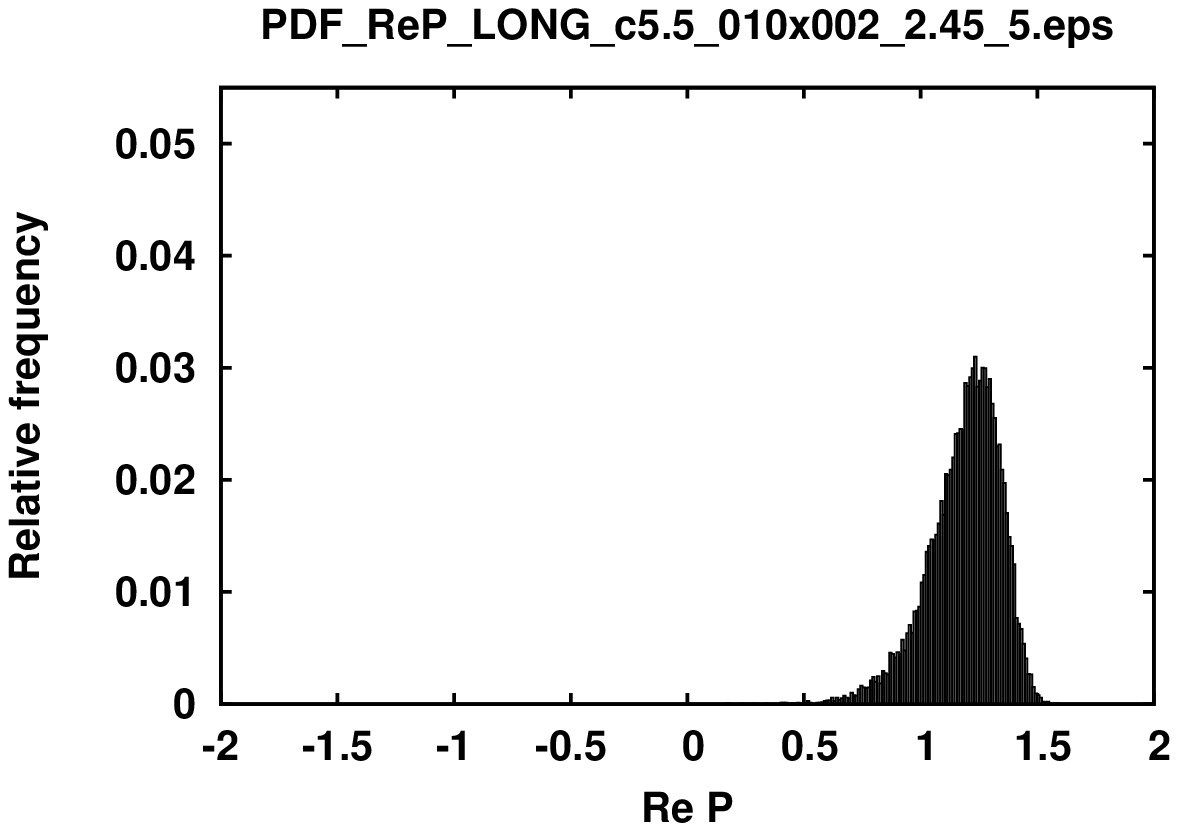}
	\includegraphics[width=0.44\textwidth]{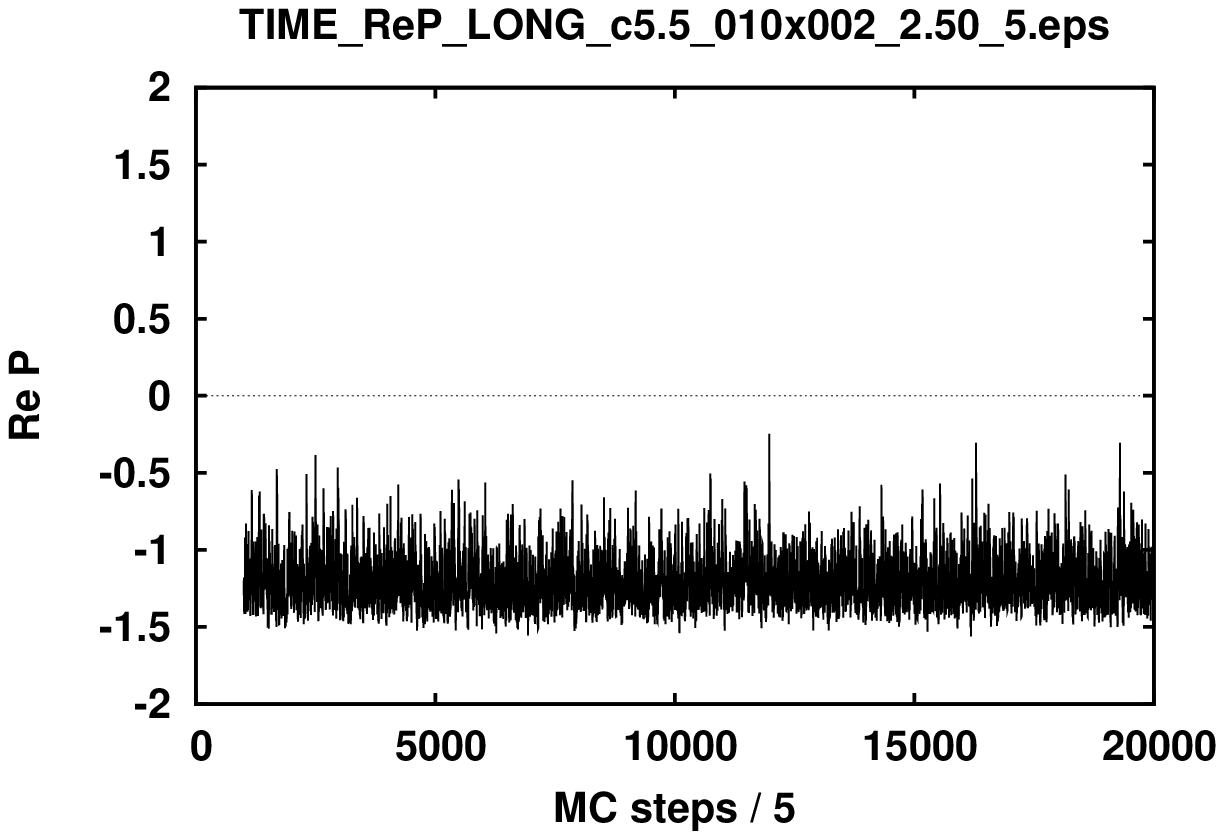} \hspace{4mm} \includegraphics[width=0.44\textwidth]{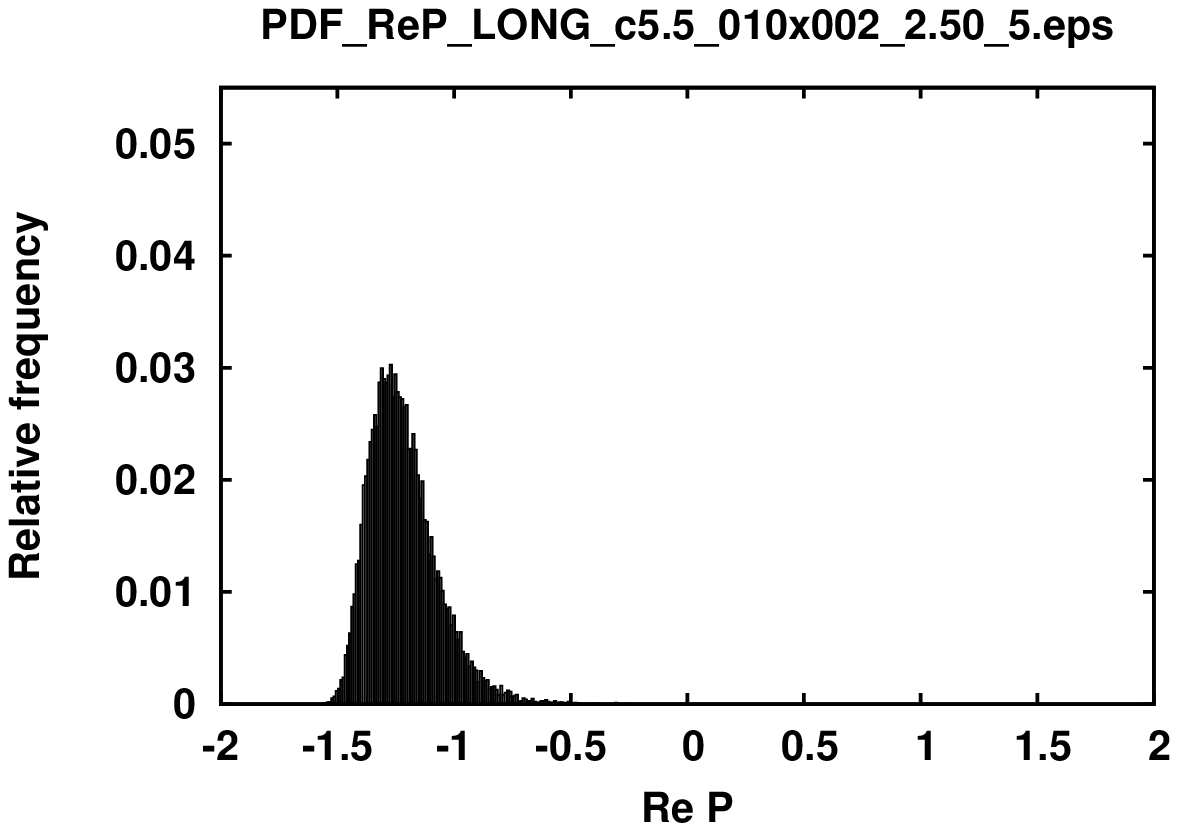}
	\includegraphics[width=0.44\textwidth]{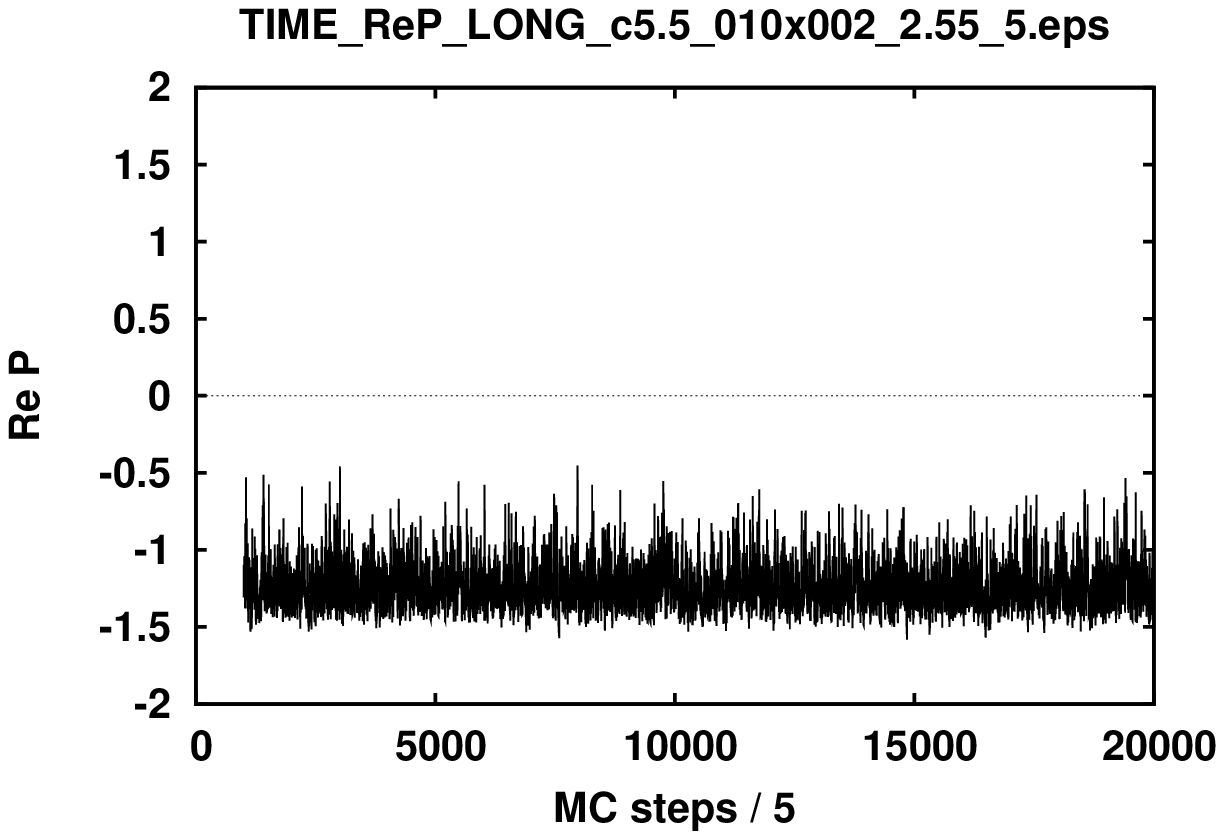} \hspace{4mm} \includegraphics[width=0.44\textwidth]{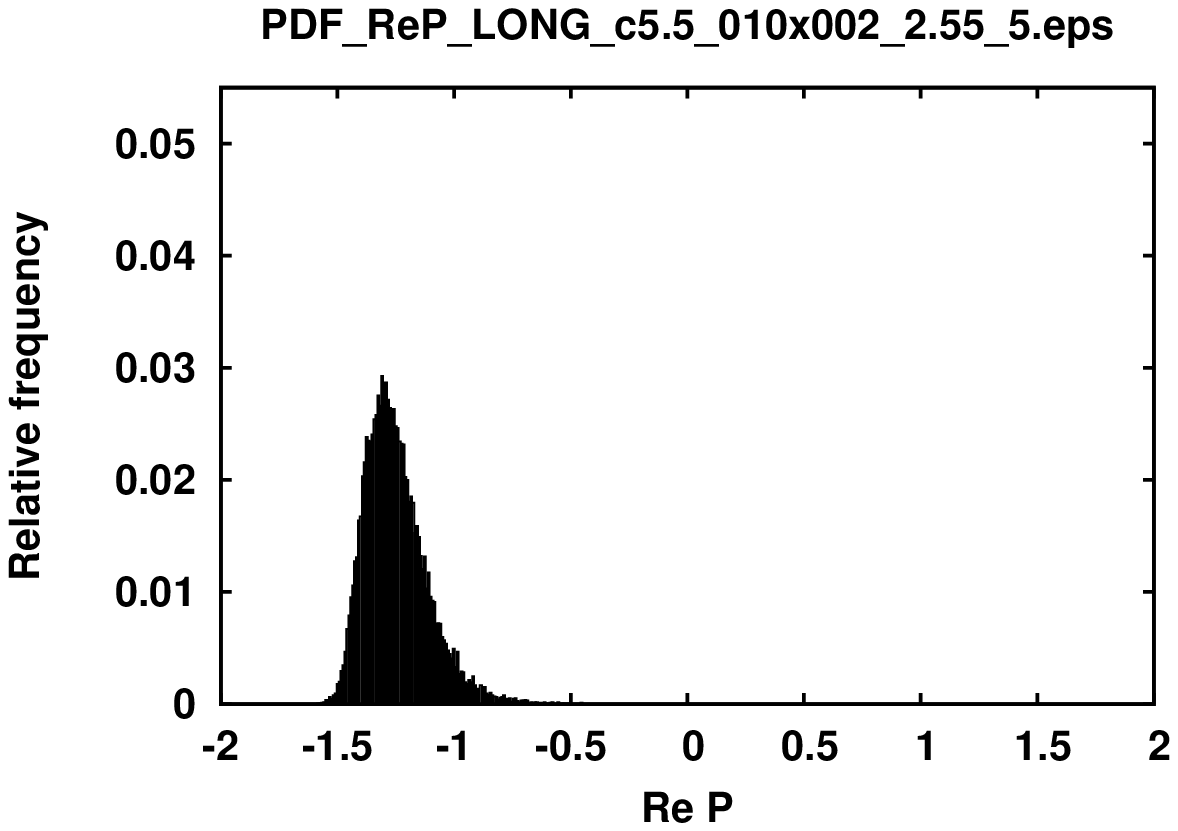}
\end{center}
\caption{ \label{FigReP_2.40_2.55}
	The Monte Carlo evolution and distribution of $\rep$ for the
	couplings $4/g^2 = 2.40, 2.45, 2.50$ and $2.55$ using $c=5.5$ from the top to the bottom.
	For these couplings only one symmetry breaking maximum occurs representing a well-developed
	deconfinement phase.
}

\end{figure}

Similar pictures from Monte Carlo simulations with fluctuating inverse temperature using the parameter
$c=5.5$ are plotted in the figures \ref{FigReP_1.80_1.95} -- \ref{FigReP_2.40_2.55}. Here the effect of
the width in the possible temperature values is clearly seen in the larger fluctuations of the
order parameter compared to the reference case $c=1024.0$ at the same coupling.
Also the critical inverse coupling strength moves towards higher values for $c=5.5$.
In Fig.\ref{FigReP_2.10_2.15} we zoom to the neighborhood of the critical coupling:
The distribution of the $\rep$ values are characteristically wide. In the third row, at
$4/g^2=2.14$, the distribution of possible values is almost flat between $-1$ and $1$.
(Due to the $SU(2)$ trace normalization, as we use it, the maximal absolute value of the order parameter is $2$.)
The intermittent behavior between positive and negative values of $\rep$, a sure sign of the
restoration of the center symmetry $Z_2$, can be catched until the value $4/g^2=2.20$, as it
can be inspected in Fig.\ref{FigReP_2.16_2.25}. For even higher inverse coupling strength the observational
sample is too short to observe this effect.


\begin{figure}

\includegraphics[width=0.7\textwidth,angle=-90]{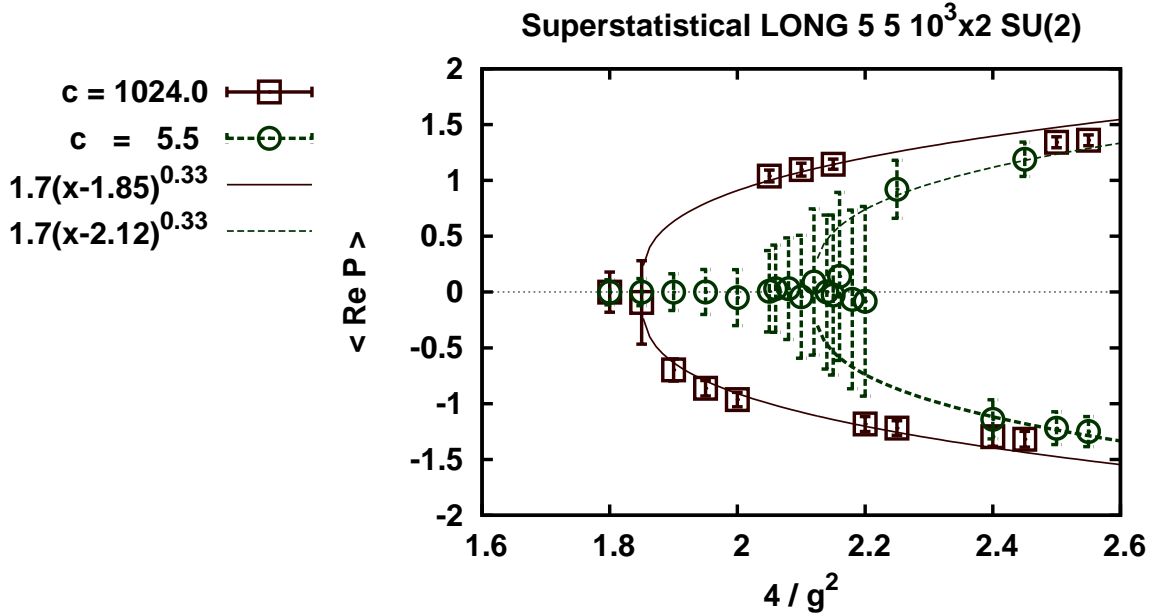}

\caption{ \label{FIG11}
  Results on Polyakov Loop spatial average
  expectation values in long runs (100.000 Monte Carlo steps, each 5-th kept)
  on $10^3\times 2$ lattices at $c=5.5$
  (red squares) and at $c=1024.0$ (green circles). The Gaussian widths are indicated
  by error bars. The transition point, i.e. the critical coupling strenth, $x=4/g^2_c$,
  is estimated by a functional fit, $\rep \sim (x-x_c)^{1/3}$. 
}

\end{figure}

How to estimate the critical coupling for the  appearence of the nonzero order parameter?
The method closest to the traditional one\cite{KUTIETAL} is to take the average value over the statistics.
In Fig.\ref{FIG11} we plot $\exv{\rep}$ over the longer Monte Carlo runs presented above
with their distribution. There is a characteristic difference between the $c=5.5$ and the $c=1024.0$
cases. A possible fit to the average values is given by a fractional power; it seems that a $1/3$
power-law behavior describes the critical scaling well. Of course, on the basis of the present data
a square root behaviour also cannot be excluded. The obtained positions of the critical couplings
differ: $4/g_c^2 \approx 1.85$ for $c=1024.0 $ while $4/g_c^2 \approx 2.12$ for $c=5.5$.


\begin{figure}

\includegraphics[width=0.7\textwidth,angle=-90]{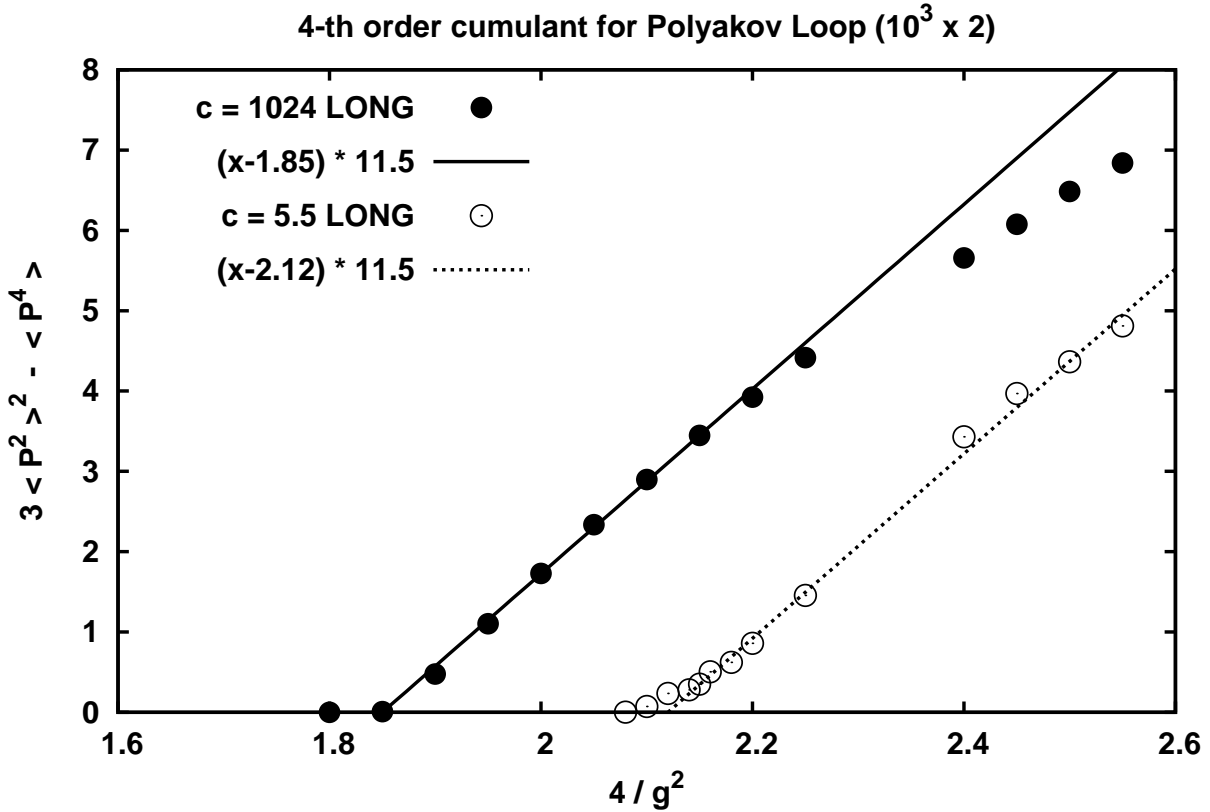}

\caption{ \label{FIGCumul}
  Fourth order cumulants of the Polyakov Loop spatial average
  in long runs (100.000 Monte Carlo steps, each 5-th kept)
  on $10^3\times 2$ lattices at $c=5.5$
  (full circles) and at $c=1024.0$ (open circles). 
  The critical coupling strenth, $x=4/g^2_c$,
  is obtained by a linear fit to the smaller nonzero values. 
}

\end{figure}

\begin{figure}

\includegraphics[width=0.7\textwidth,angle=-90]{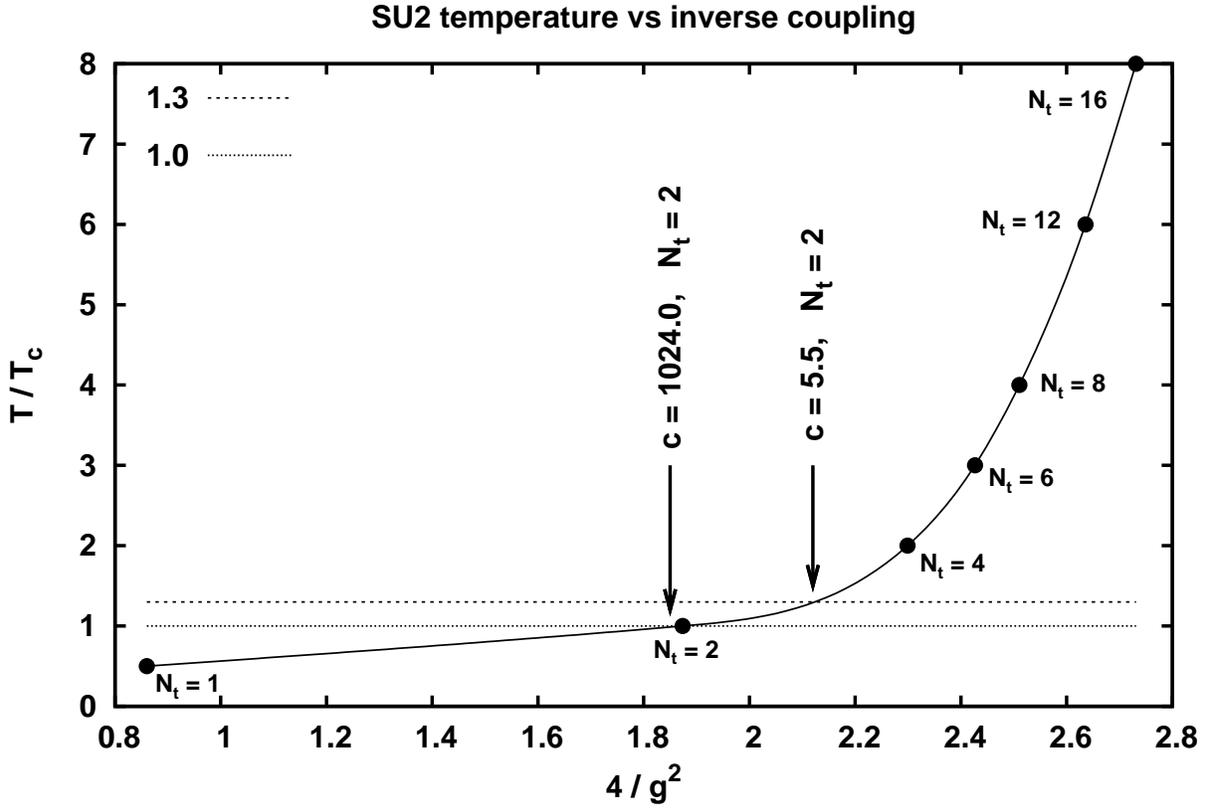}

\caption{ \label{FIG12}
  Deconfinement temperatures based on \cite{Velytsky} vs inverse coupling strength obtained from
  different size lattice simulations ($N_t$ values are indicated on the plot). 
  The arrows point to our findings of critical temperatures with $N_t=2$ for $c=1024.0$ and $c=5.5$, respectively. 
  The corresponding horizontal lines are drawn at $1.00$ and $1.30$ with respect to the $c=\infty$ case.
}

\end{figure}

For drawing conlcusions relevant to the physics the inverse lattice couplings
have to be related to temperatures. Figure \ref{FIG12} presents $T/T_c$ ratios 
versus the inverse coupling, $4/g^2$ for $N_t=2$ lattices, based
on data for critical couplings on different $N_t$-sized lattices \cite{Velytsky}. 
Although those simulations were carried out without
temperature fluctuations, i.e. taking $c=\infty$, we use them as a first estimate
for the temperature -- coupling correspondence. The critical coupling in our calculation
for $c=1024.0$ is close to the result obtained previously on same sized ($N_t=2$) lattices.
The critical coupling at $c=5.5$ -- following the $c=\infty$ line of constant physics --
corresponds on the other hand to a temperature which is $1.3$ times higher than the usual value.


\section{Conclusion}

\begin{enumerate}
\item For $c=5.5$ (a realistic value from $p_T$ spectra) the critical coupling at the deconfinement phase transition
shifts towards higher values. To this value an increase of the deconfinement temperature is
obtained at $T_c(5.5)\approx 1.3T_c(1024) \approx 1.3T_c(c=\infty)$.

\item Aiming at the same $1/T$ value for the simulation, i.e. $\exv{\theta}=1$, the temperature
is expected to make an increase of about $20$ per cent due to $\exv{1/\theta}=c/(c-1)\approx 1.22$.
This shows the same trend as obtained by the Monte Carlo simulations, but not its whole magnitude.

\item We obtained, assuming the traditional scaling dependence between coupling and physical temperature,
an increase of $15$ per cent in $4/g^2_c$ leading to about an increase of $30$ per cent in $T_c$.
The dynamical effect is definitely larger than the trivial statistical factor of $1.22$.

\item Therefore experiments aiming at producing quark matter under circumstances characteristric
to high energy collisions should  consider the possibility of an about $30$ per cent higher $T_c$
then predicted by traditional Monte Carlo lattice calculations. A possible measurement of the
value of the width parameter $c$ can be achieved by analyzing event-by-event spectra.
\end{enumerate}

These preliminary conclusions are based on a comparison with the $c=\infty$ traditional results.
In future works we aim to explore the $T/T_c - 4/g^2$ curve and possibly the renormalization
of physical quantities under the condition of fluctuating temperature with finite $c$ values.

\section*{Acknowledgment}

This work has been supported by the Hungarian National Science Fund, 
OTKA (K68108) and by the T\'AMOP 4.2.1/B-09/1/KONV-2010-0007  project co-financed
by the European Union and the European Social Fund.
Partial support from the Helmholtz International Center (HIC) for FAIR
within framework of the Landes-Offensive zur Entwicklung Wirtschaftlich-\"Okonomischer
Exzellenz (LOEWE) launched by the State of Hesse, Germany.
Discussions with Prof. B. M\"uller and A.Jakov\'ac are gratefully acknowledged.


\end{document}